\documentclass[usenatbib]{mnras}
\usepackage{mathptmx}
\usepackage{color}
\usepackage{xcolor}
\usepackage{graphics,graphicx,amssymb,amsmath}
\usepackage{subfigure}
\usepackage{multirow}
\usepackage{soul}
\soulregister\citep7
\soulregister\citet7
\soulregister\ref7

\newcommand{\tabincell}[2]{\begin{tabular}{@{}#1@{}}#2\end{tabular}}

\def\xs{30 Doradus}

\def\suzaku{{\sl Suzaku}}
\def\chandra{{\sl Chandra}}

\def\xmm{{\sl XMM-Newton}}

\date{Accepted XXX. Received YYY; in original form ZZZ}

\pubyear{2020}

\begin{document}
\title[]{X-ray spectroscopy of the starburst feedback in 30 Doradus}

\author[]{Yingjie Cheng,  Q. Daniel Wang\thanks{Contact e-mail:wqd@umass.edu},  \& Seunghwan Lim\\
  Department of Astronomy, University of Massachusetts,   Amherst, MA 01003, USA\\}

\maketitle
\begin{abstract}
X-ray observations provide a potentially powerful tool to study starburst feedback. The analysis and interpretation of such observations remain challenging, however, due to various complications, including the non-isothermality of the diffuse hot plasma and the inhomogeneity of the foreground absorption. We here illustrate such complications and a way to mitigate their effects by presenting an X-ray spectroscopy of the 30 Doradus  nebula in the Large Magellanic Clouds, based on a 100 ks {\sl Suzaku} observation. We measure the thermal and chemical properties of the hot plasma and quantitatively confront them with the feedback expected from embedded massive stars. We find that our spatially resolved measurements can be well reproduced by a global modeling of the nebula with a log-normal temperature distribution of the plasma emission measure and a log-normal foreground absorption distribution. The metal abundances and total mass of the plasma are consistent with the chemically enriched mass ejection expected from the central OB association and a $\sim 55\%$ mass-loading from the ambient medium. The total thermal energy of the plasma is smaller than what is expected from a simple superbubble model, demonstrating that important channels of energy loss are not accounted for. Our analysis indeed shows tentative evidence for a diffuse non-thermal X-ray component, indicating that cosmic-ray acceleration needs to be considered in such a young starburst region. Finally, we suggest that the log-normal modeling may be suitable for the X-ray spectral analysis of other giant HII regions, especially when spatially resolved spectroscopy is not practical. 
\end{abstract}

\section{INTRODUCTION}
\label{s:intro}
In the interstellar medium (ISM), feedback from massive stars is believed to play an essential role in regulating the evolution of their host galaxies. This feedback can be in various forms such as radiation pressure, stellar winds, and supernovae (SNe). How the feedback actually affects the ISM remains poorly understood \citep[e.g.][]{krause14}. Nevertheless, theoretical recipes \citep[e.g.][]{agertz13} have been developed to account for these effects in numerical simulations of galaxy formation and evolution. Different recipes can result in vastly different outcomes of such simulations. Observationally, it has long been difficult to quantify the effects of the feedback on its immediate surroundings, let alone its collective impact on the global ISM. As a result, the models or recipes have hardly been tested, observationally.  

\begin{figure*}
\includegraphics[width=1\linewidth]{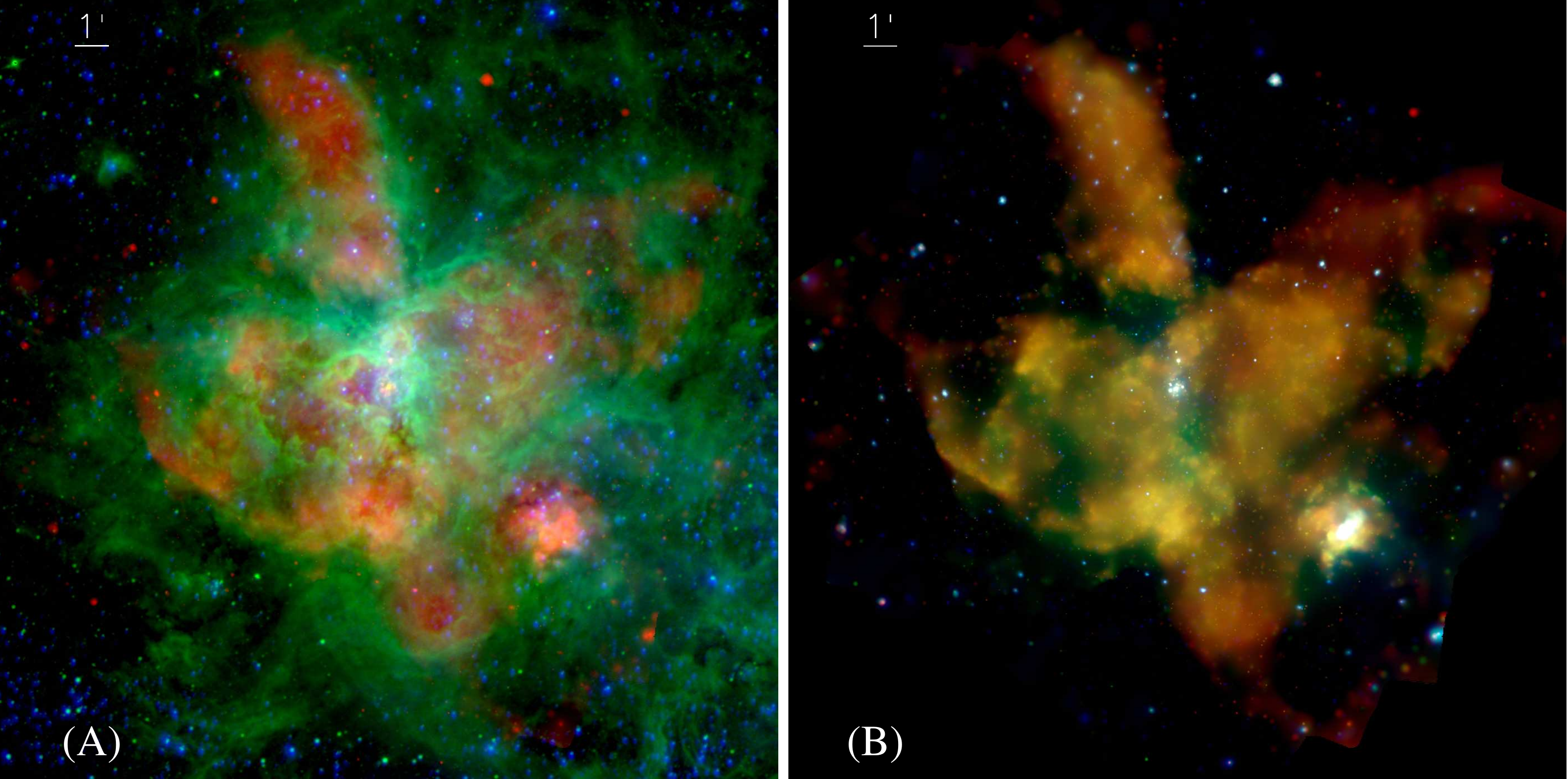}
\caption{Multi-wavelength montages of \xs: (A) 0.5-2~keV (red), H$\alpha$ (green), and UV (blue); (B) \chandra\ images in the 0.5-1~keV (red), 1-2~keV (green), and 2-8~keV (blue) bands. The H$\alpha$ image was taken with the 4m telescope at Cerro Tololo Inter-American Observatory \citep{Chu}, while the UV image was from the GALEX survey.}
\label{f:chandra}
\end{figure*}

X-ray observations of individual massive starburst regions are potentially a powerful tool to probe the effects of the stellar feedback on the ISM. Indeed, such observations, especially those made with modern X-ray instruments on-board observatories such as \chandra\ and \xmm, have shown that diffuse X-ray emission is ubiquitous in such regions, although its nature remains quite uncertain \citep[e.g.,][]{Strickland, Richings}. Probably the best laboratory to conduct X-ray studies of starburst regions at present  is \xs\ in the Large Magellanic Cloud (LMC; D=50 kpc). Because of its proximity and minimal line-of-sight confusion, \xs\ has been intensively studied in all wavelength bands as a Rosetta stone for understanding starbursts.
Recent star formation in the nebula is clearly dominated by a central burst  over the past several Myrs \citep{selman99, grebel00, rubio98, walborn99, cignoni15}. Both the initial mass function of the enclosed massive stars and the metallicity of the ambient HII medium are also well determined \citep[e.g.,][]{mathis85, Lopez2011, Pellegrini11}. 
The nebula, extending $\sim 10^2$~pc outward from the central R136 cluster, embedded within an OB association NGC~2070, is driven mainly by the high pressure of the enclosed hot gas (Fig.~\ref{f:chandra}; \S~\ref{s:res}; \citealt{Wang2, wang99, Townsley1, Pellegrini10}).

As demonstrated in Fig.~\ref{f:chandra}, \xs\ is a complex of many blisters and bubbles filled with hot plasma, traced by diffuse soft X-ray emission. The spectra of the emission show distinct lines and thus much of the emission must be thermal (due to collisional excitation in diffuse hot plasma; e.g. \citealt{Townsley1}). 

However, even in this well-studied case, many aspects of the emission still need to be carefully examined. The emission could be contaminated significantly by various other X-ray-emitting processes \citep[e.g.,][]{Persic2002,Liu2012,Zhang2014}. 
In particular, non-thermal emission due to the synchrotron and/or inverse-Compton scattering could be important, especially at relatively high photon energies ($\gtrsim 2$~keV), as has been suggested for several other star-forming regions (e.g., \citealt{Ezoe}; \S~\ref{ss:ntx}). The decomposition of such components has been difficult though, due to the limited sensitivity, spectral resolution and energy coverage of the X-ray data in existing studies. This difficulty limits the possibility to reliably measure the thermal and chemical properties of the plasma, let alone the effect of the stellar feedback on the evolution of the nebula.
Physically, the temperature (distribution) of the plasma depends on the accumulative efficiency of various cooling processes, direct (via. radiation) or indirect (via. mass-loading from the surrounding ISM). The mass-loading from cool gas also dilutes the chemical enrichment from the stellar ejection. These processes are difficult to model, especially at the early evolutionary stage of a starburst region. 

In principle, we may directly measure accumulated metal yields of massive stars, as well as the mass of the hot plasma in such a starburst region as \xs. Over a period of several $10^6$ yrs, the plasma, enriched by the mass ejection from stellar winds and SNe, is typically still well enclosed by a dense cool ambient medium, forming a structure often (simplistically) called a superbubble or a supershell. Such a superbubble is a closed box, as far as the enrichment is concerned. Also, the collisional ionization equilibrium (CIE) should be a good approximation for the enclosed plasma because of its age. In contrast, gas in a young ($\lesssim 10^4$ yrs) supernova remnant (SNR) typically retains the memory of a complex ionization history, which cannot be easily modeled. Furthermore, the emission from the CIE plasma in a superbubble at a temperature of a few times $10^6$ K is dominated by lines, tracing key $\alpha$-elements such as O, Ne, Mg, and Si, as expected from the massive star enrichment. With reasonably high-resolution spectral data, it is relatively straightforward to measure the abundances of such elements. The results should hardly depend on the dynamics of the superbubble, as long as the metals from the stars are retained in the plasma. By comparing the measurements with the expected metal enrichment from stars, the mass-loading from the ISM could be inferred. 

Although the measurement methodology and the emission physics are generally simple, complications do exist in practice. As has been shown in \xs\ \citep[e.g.][]{Townsley1}, both the hot gas temperature and the foreground X-ray-absorbing gas column density can be highly nonuniform across a nebula. 
It is not yet clear as to how such non-uniformity could be properly handled, especially when spatially-resolved X-ray spectroscopy is not practical (e.g., for a distant starburst region). 

In this paper, we present a systematic X-ray study of \xs, based primarily on a 100 ks \suzaku\ observation, complemented by an analysis of archival data from \chandra. \textit{Suzaku} provided a spectral resolution of $\sim$100-150 eV over an energy range of 0.4-10~keV, which is a factor of $\sim$ 2 better than the effective resolution of similar (non-grating) data from \chandra\ or \textit{XMM-Newton}. With the high spectral resolution and high-energy photon sensitivity, along with a low and stable instrumental background, this observation allows for an unprecedented X-ray spectroscopic capability of the hot plasma in the nebula. 

The SNR N157B to the southwest of the nebula (Fig.~\ref{f:im_reg}) is not included in the present study. The X-ray emission of the remnant is primarily non-thermal, due to the presence of an energetic pulsar of 16-millisecond period and its associated pulsar wind nebula (PWN). Detailed spatially resolved X-ray studies have been carried out previously~\citep[e.g.][]{Chen2006}. 

The rest of this paper is organized as follows: In \S~\ref{s:obs}, the observation and step-by-step data processing procedure are described. Our main results are given in \S~\ref{s:res}. In \S~\ref{s:dis}, we address such issues as whether or not there is evidence for a non-thermal X-ray (NTX) component, how the metal abundance measurements are compared to the
expected enrichment by the massive stars in NGC~2070, how efficient the mass loading is from the ISM to the hot plasma, how the energetics of the nebula is compared with the accumulated mechanical energy input, and what X-ray spectral modeling tools might be suitable for the analysis of distant starburst regions. Our results and conclusions are summarized in \S~\ref{s:sum}. Throughout the paper, the distance to \xs\ is assumed to be 50 kpc and error bars of our measurements are at the 90\% confidence. The adopted solar-abundance table is given by \citet{Anders}, which is the default setting of APEC plasma models \citep{Smith}. 

\section{Observation, data reduction, and spectral model implementation}
\label{s:obs}

Our \textit{Suzaku} data were collected with the  X-ray Imaging Spectrometer (XIS) during a 102~ks observation (ID: 806052010) taken on Nov. 30th, 2011. The XIS consisted of four CCD cameras, each combined with an X-ray telescope. While three of them were front-illuminated (XIS0, XIS2 and XIS3), one (XIS1) was back-illuminated. The front-illuminated CCDs had relatively lower and stable non-X-ray background (NXB), while the back-illuminated one had a higher sensitivity at the lower energy band. Due to the loss of XIS2 on Nov. 9th, 2006, only the data from XIS0, XIS1, and XIS3 were collected for our observation, with a field of view of $\sim~18^{'}\times18^{'}$. 

\begin{figure*}
\includegraphics[width=1\linewidth]{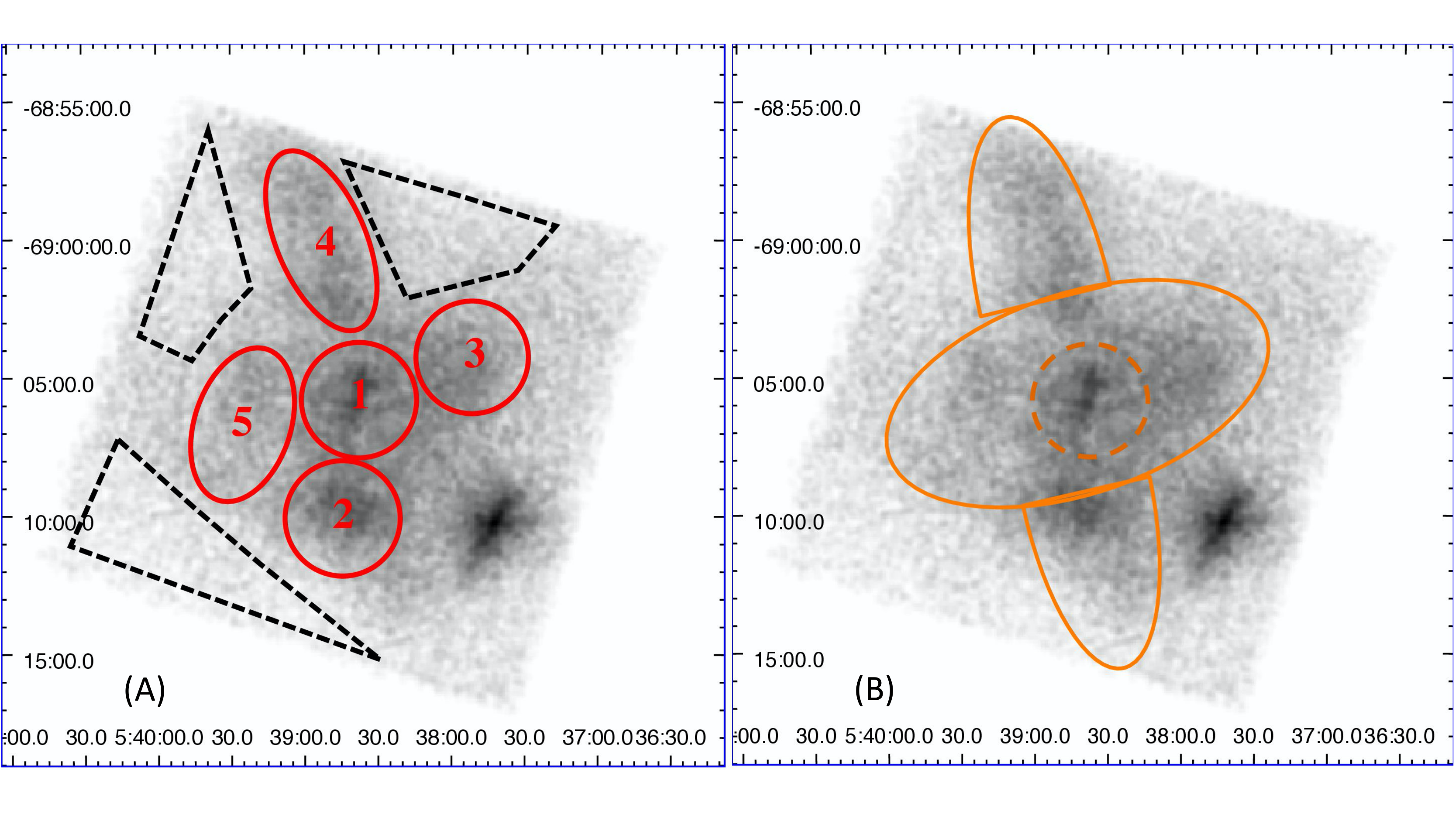}
\caption{\textit{Suzaku} intensity images of \xs\ in the $0.5-1.5$~keV band. The instrument background has been subtracted and the regions adopted for spectral extraction are outlined: (A) five individual on-nebula regions (outlined in red) and three off-nebula fields selected for background modeling (black dashed lines); (B) the entire nebula region, with or without the exclusion of Region 1 (the dashed circle) in the left panel.}
\label{f:im_reg}
\end{figure*}

We processed the \textit{Suzaku} data, using the software package HEAsoft (version 6.22) with the most up-to-date calibration database. Specifically, using the tool \textit{xisrepro} with the Version 2 standard default pipeline criteria\footnote{\url{http://heasarc.gsfc.nasa.gov/docs/suzaku/analysis/abc/}}, we screened the data and removed hot/flickering pixels. The non-X-ray background (NXB) of each CCD was generated and subtracted using the tool \textit{xisnxbgen}, while the telescope vignetting effect was corrected using the exposure map from \textit{xissim} (at 1 and 3~keV). The total effective exposure of the processed data is $\sim$101 ks. The XIS images in the 0.5-1.5~keV band is  shown in Fig.~\ref{f:im_reg}.  
 
Our spectral analysis needs to account for the instrument sensitivity and response variation across the detector field of view. As shown in Fig.~\ref{f:im_reg}A, a central region (Region 1) and four off-center regions (Regions 2, 3, 4, 5) are selected. Region 1 centers at R.A., Dec. (J2000) $= 05^h 38^m 36^s, -69^{\circ}06^{\prime}22^{\prime\prime}$, covering the central OB association NGC~2070. The sizes of all the individual regions, as included in Table \ref{t:para_derived}, will also be used later  to estimate their volumes. For each region and each of the three XIS instruments (XIS0, XIS1 and XIS3), we extract a spectrum, its corresponding non-X-ray background spectrum, effective area (arf) and spectral matrix (rmf) files, using the tools \textit{xselect}, \textit{xisnxbgen}, \textit{xissimarfgen} and \textit{xisrmfgen} respectively. We use \textit{addascaspec} to combine spectra from the two front-illuminated CCDs (XIS0 and XIS3) and their associated files, region by region. The local X-ray background is extracted from the three off-nebula fields that show no significant diffuse X-ray emission enhancement (Fig.~\ref{f:im_reg}A). As described in Appendix~\ref{a:spec_back}, we 1) characterize the intrinsic background spectrum with a model consisting of a two-temperature APEC plasma plus a power law component; 2) scale the best-fit model to predict the X-ray background spectral contribution to each on-nebula spectrum, accounting for the differences in the field coverage and instrument response; and 3) combine the contribution with the non-X-ray background to form the total background spectral contribution. The on-nebula spectra are re-binned to achieve a minimum background-subtracted S/N ratio of $3.0$. This approach simplifies the background subtraction process and makes the modeling of the on-nebula spectra much simpler, without having to deal with the background components during fitting. The plots of the spectral fits are also much cleaner. 

\begin{figure*}
\centering
\includegraphics[width=1\linewidth, trim={50 60 60 120}]{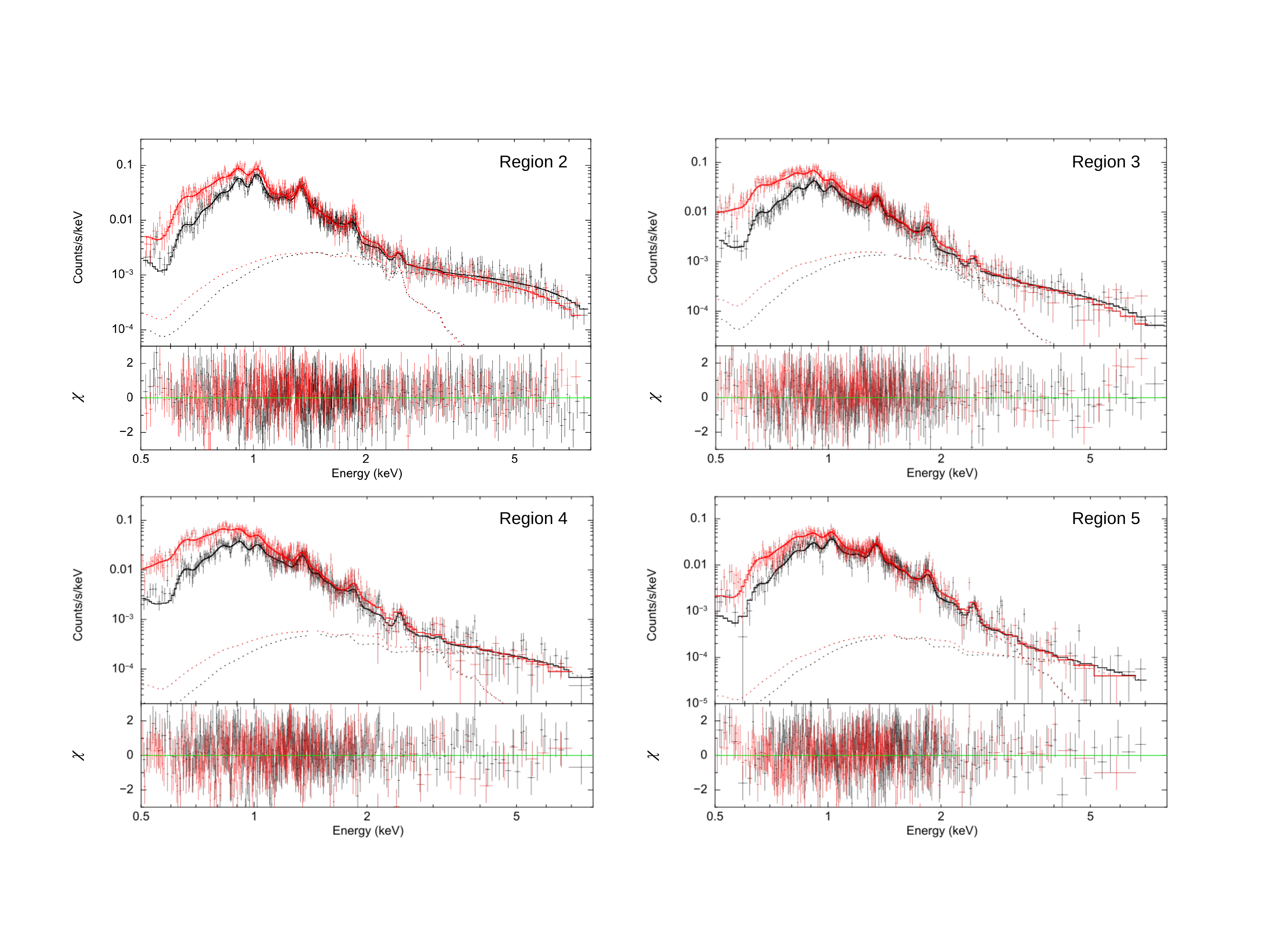}
\caption{\suzaku\ spectra of individual off-center regions, together with the corresponding best-fit spectral models (upper subpanels) and the residuals in terms of sigmas (lower subpanels). The FI (black) and BI (red) spectra are presented separately. From the bottom to the top at $\sim 1$~keV in each panel, the dotted lines of the same color represent the contributions from the power-law components and the log-normal temperature thermal plasma components with simple foreground absorption. The local X-ray backgrounds have already been subtracted along with the NXB.
}
\label{f:spec_s}
\end{figure*}


Considering the relatively poor PSF of the \suzaku\ data, we estimate its contamination effect on our spectral data extracted in various regions (S~\ref{s:res}) in Appendix~\ref{a:psf}. This estimate is focused on the spilling out of counts from NGC~2070, as well as a couple of relatively bright discrete sources in the vicinity. The contamination is most important in the hard band, potentially affecting the spectral characterization of the emission from \xs\ (see also \S~\ref{ss:ntx}).

To assist the interpretation of the \suzaku\ data, we have also reduced complementary data on \xs\ from \chandra\   X-ray Observatory. These data are obtained from the 
\chandra\ archive and are reduced with the standard CIAO software processing pipeline. The \chandra\ data include all 56 ACIS-I observations of a total exposure 2.07 Ms, which are available by June, 2020. We use the data to gain a high spatial resolution view of \xs\ and to tightly constrain the emission from discrete X-ray sources, which are detected to a flux limit of $\sim 5.8 \times 10^{31} {\rm~ergs~s^{-1}}$ in the 0.5-8~keV band \citep{wang04}. The brightest X-ray source in the field is the SNR, N157B, which contains the 16 ms pulsar and its associated wind nebula. Other sources are primarily associated with NGC~2070.

We use  the \chandra\ data here primarily to constrain the integrated spectral properties of the sources associated with NGC~2070. We have separately extracted spectra for the sources and the diffuse (source-removed) emission within $1^\prime$ of the central cluster R136 of NGC~2070. To obtain the diffuse emission, we remove a region of twice the 70\% energy-encircled radius of the PSF around each source; for a bright source ($>~0.01 \rm~counts~s^{-1}$), the radius is logarithmically scaled up accordingly to minimize the effect of the extended wing of the PSF \citep{wang04}. 
The spectrum of the diffuse emission is used as the local background in the analysis of the  source spectrum.

Finally, to characterize the variations in both  the absorbing gas column density and the plasma temperature, we have implemented the corresponding log-normal distribution models ({\sl lnabs} and {\sl vlntd}) for spectral analysis. We describe these implementations in Appendices~\ref{a:impl-lnnh} and ~\ref{a:impl} and the inference of physical properties of the plasma drawn from the log-normal temperature model is described in Appendix~\ref{a:para}. The continuous distribution of the temperature should typically be more physical than the commonly-used two-temperature plasma model. The log-normal distribution is  expected from the central limit theorem for a variable characterizing the accumulation of many independent multiplicative random processes. The plasma temperature resulting from repeated shock-heating is such a variable.
In fact, the log-normal distribution is shown to characterize the temperature structure of hot plasma quite well in various hydrodynamic simulations (e.g., \citealt{Frank}), and is also apparent in existing observational measurements \citep[e.g.,][]{Nakashima2018}. 
Regardless of the exact state of the plasma, we here consider the log-normal distribution with only one additional parameter (the dispersion of the temperature in logarithm; in contrast to at least one more parameter in plasma modeling with multiple independent temperatures) as a simple extension of the isothermal model to characterize the non-isothermality. Furthermore, as shown in Appendix~\ref{a:impl},  the isothermal or Gaussian temperature model just represents a special case of the log-normal distribution when the dispersion is zero or small. Similarly, the distribution of column densities in a turbulent non-starforming dense gas has been shown to follow an approximately log-normal form \citep{Ostriker}. Our implemented multiplicative model {\sl lnabs} (see Appendix~\ref{a:impl-lnnh}) is simply a sum of the foreground X-ray absorption weighted by a log-normal distribution of the absorbing gas column density. This model may thus be applicable to a region that is large enough to sample a range of foreground X-ray absorption variation.

\section{Analysis and Results}
\label{s:res}

Although the present study is focused on the \suzaku\  spectroscopy, we show \chandra\ X-ray images of \xs\ in Fig.~\ref{f:chandra} as a high spatial resolution reference, together with a multi-wavelength comparison.
Fig.~\ref{f:im_reg} presents the XIS images in the 0.5-1.5~keV band and illustrates various spectral regions that are used for our spectral analysis. 

The five independent on-nebula regions in Fig.~\ref{f:im_reg}A cover the most distinct blisters or bubbles in \xs, based on existing studies (e.g., \citealt{Wang2};  see also Fig.~\ref{f:chandra}). The properties (mainly temperature and metal abundances of the hot plasma, as well as the foreground absorption) of each of these five regions should be relatively uniform and could be characterized with a  simple spectral model. The results from the spectral analysis of the regions can then be used to test the modeling approach for the spectra extracted from the entire off-center region and the entire nebula, as defined in Fig.~\ref{f:im_reg}B. To simplify the estimate of the overall volume, a full ellipse and two elliptical caps are selected to enclose the essentially entire X-ray-emitting nebula.  

We take steps to achieve our analysis objectives, using the X-ray spectral fitting program XSPEC (Version 12.11.0). The X-ray spectral properties of \xs\ vary strongly from one region to another. The central region, in particular, is distinctly different. Its spectrum includes a large contribution from discrete sources and can hardly be modeled with a combination of simply components such as a single temperature (1-T hereafter) thermal plasma plus a power law, multiplied by a foreground absorption. Therefore, we choose to deal with the surrounding regions first, then the central region, and finally the integrated spectra of \xs. 

\subsection{Individual off-center regions}\label{ss:res-ind}

We first fit each spectrum of Regions 2, 3, 4 and 5 with a 1-T plasma plus power law model multiplied by a simple foreground absorption. The fits are acceptable, considering the high signal-to-noise ratio of the spectra. Nevertheless, to characterize the degree of the non-isothermality and its effect on the spectral parameter estimates, we replace the 1-T plasma model with the log-normal temperature distribution plasma model {\it vlntd}  (Appendix~\ref{a:impl}). Fig.~\ref{f:spec_s} presents the best fits to the \suzaku\ spectra in this log-normal temperature modeling. For these regions, changing the simple foreground absorption to a log-normal column density distribution absorption ({\it lnabs}) does not significantly improve the fit, though, indicating that the foreground absorption is reasonably uniform across each of these four regions. Thus, the corresponding results of the log-normal absorption are not included in the table. 

Table \ref{t:spec_indi} lists the best-fit parameters for both the single- and log-normal-temperature modeling cases. 
The 90\% confidence ranges of all the parameters are calculated by the Markov Chain Monte Carlo (MCMC) method implemented in XSPEC. 
The comparison of the $\chi^2/dof$ values shows that the log-normal temperature distribution model describes the spectra considerably better than the 1-T model. 
As shown in the table, the temperature dispersion is significant for all the individual regions. In particular, the dispersion $\sigma_{x}$ for Region 4 reaches up to $\sim 0.74$, which is comparable to its $\bar{x}$ value of $\sim 0.84$. 
For Regions 2 and 5, the mean temperatures inferred from the best-fit $\bar{x}$  do not significantly differ from those from the 1-T plasma modeling and are significantly higher than those in Regions 3 and 4. In these latter regions, the mean temperatures from the log-normal modeling are considerably lower, while the emission measures are greatly increased by a factor of $\sim2$ than the corresponding 1-T values.
For all the individual regions, by changing to the log-normal modeling, the best-fit elemental abundances show a systematic increasing trend, and exceeds the 90\% confidence range for most cases. In general, the abundances are higher than the mean values of the ISM in the LMC ($\sim 0.5~solar$). To be more specific, when comparing with the dust-depletion corrected abundances derived from SNR radiative shocks \citep{dopita19}, our best-fit abundances of O, Ne and Mg are substantially higher while Si and Fe are slightly lower. The line features in the spectra can all be accounted for by the thermal plasma component. Furthermore, the column density differences among the regions are also considerable. 

The power law component is statistically significant in all the above and later fits and accounts for much of the high-energy tails of the spectra above $\sim 3$~keV (e.g., Fig.~\ref{f:spec_s}). This requirement for a power-law component is a bit surprising, because the discrete source contribution should be insignificant in the off-center regions. As described in Appendix~\ref{a:psf}, 
the count rates in the hard band of the off-center spectra are also much greater than the expected spilling-out from neighboring relatively bright sources due to the PSF.  Therefore, the bulk of the hard band spectra should be diffuse in origin.

\begin{table*}
\caption{Spectral analysis results of off-center nebula regions${}^{\rm a}$}
\begin{tabular}{lcccc}
\hline
Parameters & Region2 & Region3 & Region4 & Region5 \\
\hline
$N_{\rm H}\ (10^{22}\rm{cm}^{-2})$ & \tabincell{c}{$0.37(0.34-0.41)$\\$0.36(0.34-0.38)$} & \tabincell{c}{$0.30(0.26-0.34)$\\$0.36(0.34-0.38)$} & \tabincell{c}{$0.27(0.24-0.31)$\\$0.31(0.29-0.33)$} & \tabincell{c}{$0.47(0.42-0.53)$\\$0.51(0.46-0.54)$} \\

$EM\ (10^{59}\rm{cm^{-3})}$ & \tabincell{c}{$22(18-27)$\\$22(19-25)$} & \tabincell{c}{$25(18-34)$\\$56(48-65)$} & \tabincell{c}{$18(13-25)$\\$36(27-42)$} & \tabincell{c}{$22(14-30)$\\$22(18-25)$} \\

$kT_{\rm mean}\ (\rm{keV})$ & \tabincell{c}{$0.40 (0.39-0.41)$\\$0.42(0.41-0.43)$} & \tabincell{c}{$0.33(0.31-0.34)$\\$0.20(0.18-0.23)$} & \tabincell{c}{$0.35(0.33-0.36)$\\$0.20(0.16-0.23)$} & \tabincell{c}{$0.39(0.37-0.41)$\\$0.39(0.38-0.41)$} \\
$\bar{x}\ $ & -/$1.58(1.56-1.61)$ & -/$0.84(0.74-0.98)$ & -/$0.84(0.62-0.98)$ & -/$1.51(1.48-1.56)$ \\
$\sigma_x\ $ & -/$0.25(0.20-0.29)$ & -/$0.57(0.50-0.65)$ & -/$0.74(0.63-0.88)$ & -/$0.32(0.27-0.37)$ \\

O (solar)& \tabincell{c}{$0.8(0.5-1.1)$\\$0.9(0.7-1.1)$} & \tabincell{c}{$0.4(0.3-0.5)$\\$0.6(0.5-0.7)$} & \tabincell{c}{$0.5(0.4-0.6)$\\$0.7(0.6-0.9)$} & \tabincell{c}{$0.5(0.3-0.8)$\\$0.9(0.8-1.3)$} \\
Ne (solar)& \tabincell{c}{$1.2(1.0-1.6)$\\$1.8(1.6-2.1)$} & \tabincell{c}{$0.5(0.4-0.7)$\\$0.9(0.8-1.1)$} & \tabincell{c}{$0.5(0.4-0.7)$\\$0.9(0.8-1.1)$} & \tabincell{c}{$0.7(0.5-1.1)$\\$1.2(1.0-1.6)$} \\
Mg (solar)& \tabincell{c}{$0.8(0.7-1.0)$\\$1.1(1.0-1.4)$} & \tabincell{c}{$0.5(0.4-0.7)$\\$0.8(0.7-0.9)$} & \tabincell{c}{$0.4(0.3-0.6)$\\$0.8(0.7-1.1)$} & \tabincell{c}{$0.5(0.4-0.8)$\\$0.9(0.7-1.1)$} \\
Si (solar)& \tabincell{c}{$0.5(0.4-0.6)$\\$0.5(0.4-0.6)$} & \tabincell{c}{$0.6(0.5-0.8)$\\$0.5(0.4-0.7)$} & \tabincell{c}{$0.3(0.2-0.5)$\\$0.3(0.2-0.4)$} & \tabincell{c}{$0.4(0.3-0.6)$\\$0.5(0.3-0.6)$} \\ 
Fe (solar)& \tabincell{c}{$0.2(0.1-0.3)$\\$0.3(0.2-0.4)$} & \tabincell{c}{$0.2(0.1-0.3)$\\$0.4(0.3-0.5)$} & \tabincell{c}{$0.2(0.1-0.3)$\\$0.4(0.3-0.5)$} & \tabincell{c}{$0.2(0.1-0.3)$\\$0.3(0.2-0.4)$} \\

$\alpha$ (power-law index)&
\tabincell{c}{$1.43(1.30-1.55)$\\$1.26(1.12-1.38)$} & \tabincell{c}{$2.32(2.05-2.60)$\\$1.90(1.57-2.09)$} & \tabincell{c}{$2.17(1.84-2.52)$\\$1.05(0.76-1.47)$} & \tabincell{c}{$2.92(2.13-3.67)$\\$1.54(1.38-1.73)$} \\

$\rm{Norm^{pow}}$ $(10^{-3}$ $\rm{KeV^{-1}}$ $\rm{cm^{-2}}$ $\rm{s^{-1}}$)& \tabincell{c}{$1.65 (1.37-1.97)$\\$1.24(1.00-1.50)$} & \tabincell{c}{$1.77 (1.25-2.44)$\\$0.89(0.55-1.11)$} & \tabincell{c}{$1.24 (0.81-1.86)$\\$0.19(0.11-0.35)$} & \tabincell{c}{$1.50 (0.53-3.64)$\\$0.16(0.06-0.38)$} \\

$\rm{F^{pow}}$$(10^{-12}$ ergs $\rm{cm^{-2}}$ $\rm{s^{-1})}$ & \tabincell{c}{$3.45(3.32-3.58)$\\$2.90(2.79-3.02)$} & \tabincell{c}{$2.57(2.41-2.73)$\\$1.45(1.32-1.57)$} & \tabincell{c}{$2.00(1.86-2.14)$\\$0.54(0.47-0.61)$} & \tabincell{c}{$1.33(1.18-1.49)$\\$0.27(0.20-0.35)$} \\

$\rm{F^{obs}}$$(10^{-11}$ ergs $\rm{cm^{-2}}$ $\rm{s^{-1})}$ & \tabincell{c}{$2.74(2.71-2.77)$\\$2.74(2.71-2.77)$} & \tabincell{c}{$2.08(2.05-2.11)$\\$2.12(2.09-2.15)$} & \tabincell{c}{$1.85(1.82-1.88)$\\$1.89(1.87-1.92)$} & \tabincell{c}{$1.42(1.40-1.44)$\\$1.41(1.39-1.44)$} \\
$\chi^2/dof$ & $\frac{840}{889}$/$\frac{822}{888}$ & $\frac{759}{680}$/$\frac{724}{679}$ & $\frac{776}{699}$/$\frac{724}{698}$ & $\frac{764}{673}$/$\frac{746}{672}$ \\
\hline
\end{tabular}

Note:  ${}^{\rm a}$Spectral parameters from the single and log-normal temperature models are presented separately (corresponding to the first and second rows of each parameter or separated by $/$). The 90\% confidence ranges are included right after the best-fit values. For log-normal temperature distribution models, $x \equiv {\rm ln} T$, where T is the temperature in units of $10^6~K$, and $\sigma_x$ is the dispersion of $x$. The emission measure $EM = \int n_e n_H d V$ is inferred from the  \textit{apec} model normalization $\equiv \dfrac{10^{-14}}{4\pi D^2} EM$, where $D$ is the distance ($\sim$ 50 kpc), $n_e$ and $n_H$ are the electron and H densities. $\rm{F^{pow}}$ gives the observed (absorbed) flux of the power-law component while $\rm{F^{obs}}$ gives the total observed X-ray flux.
\label{t:spec_indi}
\vspace{0.2in}
\end{table*}

\subsection{Entire off-center nebula region} 
\label{ss:off-center}
\begin{figure}
\includegraphics[width=1\linewidth,trim={30 30 20 120}]{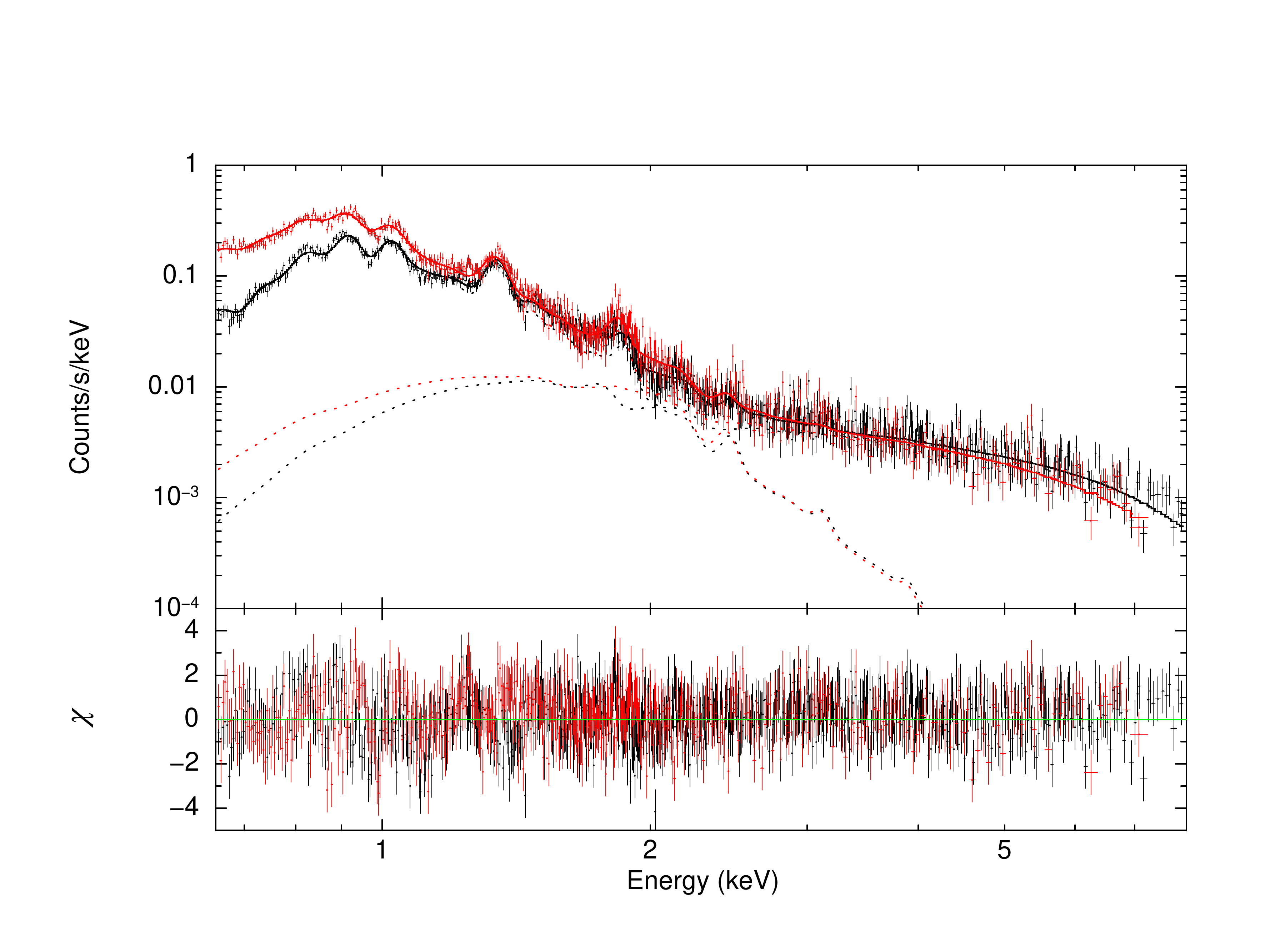}
\caption{\suzaku\  spectra of the total off-center region and the best-fit model, which comprises of a thermal plasma with a log-normal temperature distribution, a power-law component, plus a foreground absorption with a log-normal column density distribution.}
\label{f:spec_off}
\end{figure}

We next conduct spectral fits to the integrated spectra from the entire off-center region of \xs. The spectra maximize the S/N of the data, but are clearly complicated by the mixture of  plasma and foreground absorption with a range of properties. Not surprisingly, a fit with a model as used for individual regions can be firmly rejected.
We thus adopt a spectral model with the log-normal distribution for both the plasma temperature and the the absorbing gas column density (Table~\ref{t:off-center}). This modeling gives a reasonable fit to the data, judging by the reduced $\chi^2$ value, although it is still not statistically acceptable. Because of the increased S/N of the spectra, compared to those of the individual regions, systematic uncertainties (e.g., in the data calibration and background subtraction) should ideally be considered. But they are difficult to quantify. For example, our fitting in the 0.5-0.65~keV band appears to be problematic, which is at least partly due to somewhat uncertain subtraction of relatively large NXB contributions in different CCDs at such energies (Appendix~\ref{a:spec_back}).
To demonstrate this effect, Table~\ref{t:off-center} includes the results from the spectral fitting in the 0.65-8~keV range. Comparatively, the fit is significantly improved and is shown in Fig.~\ref{f:spec_off}. In this case, however, the abundance of Oxygen cannot be constrained since its line emission is now outside the energy range. Furthermore, Table~\ref{t:off-center}  further shows that  the parameters obtained here are very close to the flux-weighted means of the same parameters from fitting to individual regions (Table~\ref{t:spec_indi}). Therefore, we consider that the log-normal modeling for the temperature and absorption is a reasonable success. 

\begin{table}
\caption{Spectral analysis results of the off-center nebula as a whole }
\label{t:off-center}
\def\arraystretch{1.2}
\begin{tabular}{lccr}
\hline
Parameters & 0.65-8 keV & 0.5-8 keV & mean${}^{\rm a}$ \\
\hline
$N_{\rm H, mean} (10^{22}\rm{cm}^{-2})$ & $0.37(0.32-0.39)$ & $0.35(0.30-0.39)$ & $0.37$ \\
$\sigma_{N_{\rm H}} (10^{-2})$ & $0.1(<3)$ & $0.2(<3)$ & $/$\\

$\rm{EM(10^{59}cm^{-3})}$ & $50(46-60)$ & $82(75-96)$ & $34$\\
$kT_{\rm mean}(\rm{keV})$ & $0.28(0.27-0.29)$ & $0.20(0.18-0.22)$ & $0.31$\\
$\bar{x}\ $ &$1.18(1.14-1.21)$ & $0.84(0.74-0.94)$ &$1.20$\\
$\sigma_x\ $ & $0.50(0.44-0.56)$ & $0.70(0.65-0.76)$ & $0.46$\\
O (solar)& $/$ & $0.4(0.3-0.6)$ & $0.8$\\
Ne (solar)& $0.9(0.8-1.0)$ & $0.8(0.7-0.9)$ & $1.2$\\
Mg (solar)& $0.8(0.7-0.9)$ & $0.7(0.6-0.8)$ & $0.9$\\ 
Al (solar)& $0.7(0.6-0.8)$ & $0.7(0.6-0.9)$ & $/$\\
Si (solar)& $0.4(0.3-0.5)$ & $0.3(0.2-0.5)$ & $0.5$\\
S (solar)& $0.4(0.3-0.6)$ & $0.3(0.2-0.4)$ & $/$\\
Fe (solar)& $0.29(0.24-0.33)$ & $0.28(0.24-0.31)$ & $0.3$\\
$\alpha$ (power-law index)& $1.55(1.48-1.60)$ & $1.44(1.43-1.45)$ & $1.43$\\
\multirow{2}{1.5cm}{$\rm{Norm^{pow}}$$(10^{-3}$$\rm{keV^{-1}}$ $\rm{cm^{-2}}$ $\rm{s^{-1}}$)} & $2.0(1.9-2.2)$ & $1.7(1.6-1.8)$ & $0.7$\\
 &  &  &  \\
 \multirow{2}{2cm}{$\rm{F^{pow}}$ $(10^{-12}$ ergs $\rm{cm^{-2}}$ $\rm{s^{-1})}$} & $3.92(3.84-4.00)$ & $3.49(3.42-3.57)$ & $1.52$\\
 &  &  &  \\
\multirow{2}{2cm}{$\rm{F^{obs}}$ $(10^{-11}$ ergs $\rm{cm^{-2}}$ $\rm{s^{-1})}$} & $3.76(3.74-3.78)$ & $3.81(3.78-3.83)$ & $2.15$\\
 &  &  &  \\
$\chi^2/dof$ & $\frac{1609}{1349}=1.19$ & $\frac{1764}{1402}=1.25$ &$/$\\
\hline
\end{tabular}

Note:  ${}^{\rm a}$ For comparison, this last column lists the flux-weighted mean values of corresponding parameters over the four individual off-center regions (Table~\ref{t:spec_indi}).
\end{table}

\begin{table}
\caption{Spectral analysis results of the central region}
\label{t:spec-cen}
\begin{tabular}{lcr}
\hline
Parameters & Power+Line & Source Model \\
\hline
$N_{\rm H, mean} (10^{22}\rm{cm}^{-2})$ & $0.42(0.40-0.44)$ &$0.45(0.40-0.50)$\\
$\sigma_{N_{\rm H}} (10^{-2})$ & $0.7(0.4-1.1)$ &$2.0(1.7-2.3)$\\

$\rm{EM(10^{59}cm^{-3})}$ & $110(72-158)$&$115(77-152)$ \\
$kT_{\rm mean}(\rm{keV})$ & $0.19(0.12-0.26)$ &$0.23(0.18-0.27)$\\
$\bar{x}\ $ &$0.8(0.3-1.1)$ &$1.0(0.7-1.1)$\\
$\sigma_x$ & $0.8(0.7-1.0)$ &$0.6(0.4-0.8)$\\
O (solar)& $0.5(0.3-0.7)$ &$0.4(0.2-0.5)$\\
Ne (solar)& $0.9(0.8-1.1)$ &$0.6(0.4-0.7)$\\
Mg (solar)& $1.0(0.8-1.2)$ &$0.7(0.5-1.2)$\\ 
Al (solar)& $1.2(0.8-1.8)$ &$1.0(0.5-1.5)$\\
Si (solar)& $0.6(0.4-0.8)$ &$0.4(0.2-0.6)$\\
S (solar)& $0.7(0.5-0.9)$ &$0.2(0.1-0.3)$\\
Fe (solar)& $0.3(0.2-0.5)$ &$0.2(0.1-0.3)$\\
$\alpha$ (power-law index)& $2.56(2.50-2.62)$ &$-$\\
$\rm{Norm^{pow}}$ $(10^{-2}$$\rm{keV^{-1}}$ $\rm{cm^{-2}}$ $\rm{s^{-1}}$)& $2.0(1.5-2.5)$  &$-$\\
LineE$(\rm{keV})$ & $6.63(6.57-6.70)$ & $-$ \\
Line$\sigma(\rm{keV})$ &$0.09(<0.19)$ & $-$ \\
$\rm{Norm^{line}}(10^{-5})$ &$5.7(4.0-7.5)$ & $-$ \\
factor & $-$ & $9.7(9.4-10.0)$ \\
$\rm{F^{pow}}$ $(10^{-11}$ ergs $\rm{cm^{-2}}$ $\rm{s^{-1})}$ & $2.26(2.22-2.30)$& $-$
\\ 
$\rm{F^{obs}}$ $(10^{-11}$ ergs $\rm{cm^{-2}}$ $\rm{s^{-1})}$ & $6.94(6.88-7.00)$& $6.94(6.88-7.00)$
\\ 
$\chi^2/dof$ & $\frac{1236}{1227}=1.01$ &$\frac{1256}{1231}=1.02$\\
\hline
\end{tabular}

Note: In the first column, a power-law model and a $\sim6.6$~keV Gaussian line is added to characterize the hard-band emission and the Fe line; for the second column, a scaled \chandra\ point source model is added instead, and the best-fit scaling factor is given.

\vspace{0.5cm}
\end{table}



\subsection{Central nebular region}
\label{ss:central}

\begin{table}
\caption{\chandra\ spectral analysis results of discrete sources within $1^{\prime}$ of NGC~2070}\label{t:spec_chandra}
\begin{tabular}{lr}
\hline
Parameter & Value \\
\hline
$N_{\rm H, mean} (10^{22} {\rm~cm^{-2}})$ &$0.65(0.61-0.69)$\\
$\sigma_{N_{\rm H}} (10^{-2})$ & $0.7(0.3-1.1)$ \\
$\alpha$ (power-law index) &$2.9(2.8-3.0)$\\
$\rm{Norm^{pow}} (10^{-4} {\rm~kev^{-1}cm^{-2}s^{-1}})$ &$2.1(1.9-2.3)$\\
$\rm{F^{pow}} (10^{-13}~ergs~cm^{-2}s^{-1})$ &$1.55(1.52-1.57)$\\
$kT_{mean,1}$  (keV) &$2.8(2.6-2.9)$\\
$\bar{x}_1\ $ &$3.48(3.41-3.52)$ \\
$\sigma_{x,1}$ & $0.5(0.4-0.7)$ \\
$\rm{EM_1} (10^{59}cm^{-3})$ &$0.21(0.19-0.23)$\\
$kT_{mean,2}$ (keV) &$0.76(0.71-0.80)$\\
$\bar{x}_2\ $ &$2.18(2.11-2.23)$ \\
$\sigma_{x,2}$ & $0.3(0.2-0.5)$ \\
$\rm{EM_2} (10^{59}cm^{-3})$ &$0.22(0.19-0.26)$\\
$\chi^2/dof$ &$\frac{565}{450}=1.26$ \\
$\rm{F^{obs}} (10^{-13}~ergs~cm^{-2}s^{-1})$ &$5.67(5.64-5.69)$ \\
\hline
\end{tabular}

Note: The metal abundances are fixed at 0.5 solar, which is appropriate for the ISM in the LMC \citep{Russell}.
\end{table}

\begin{figure}
\includegraphics[width=1\linewidth,trim={30 30 20 120}]{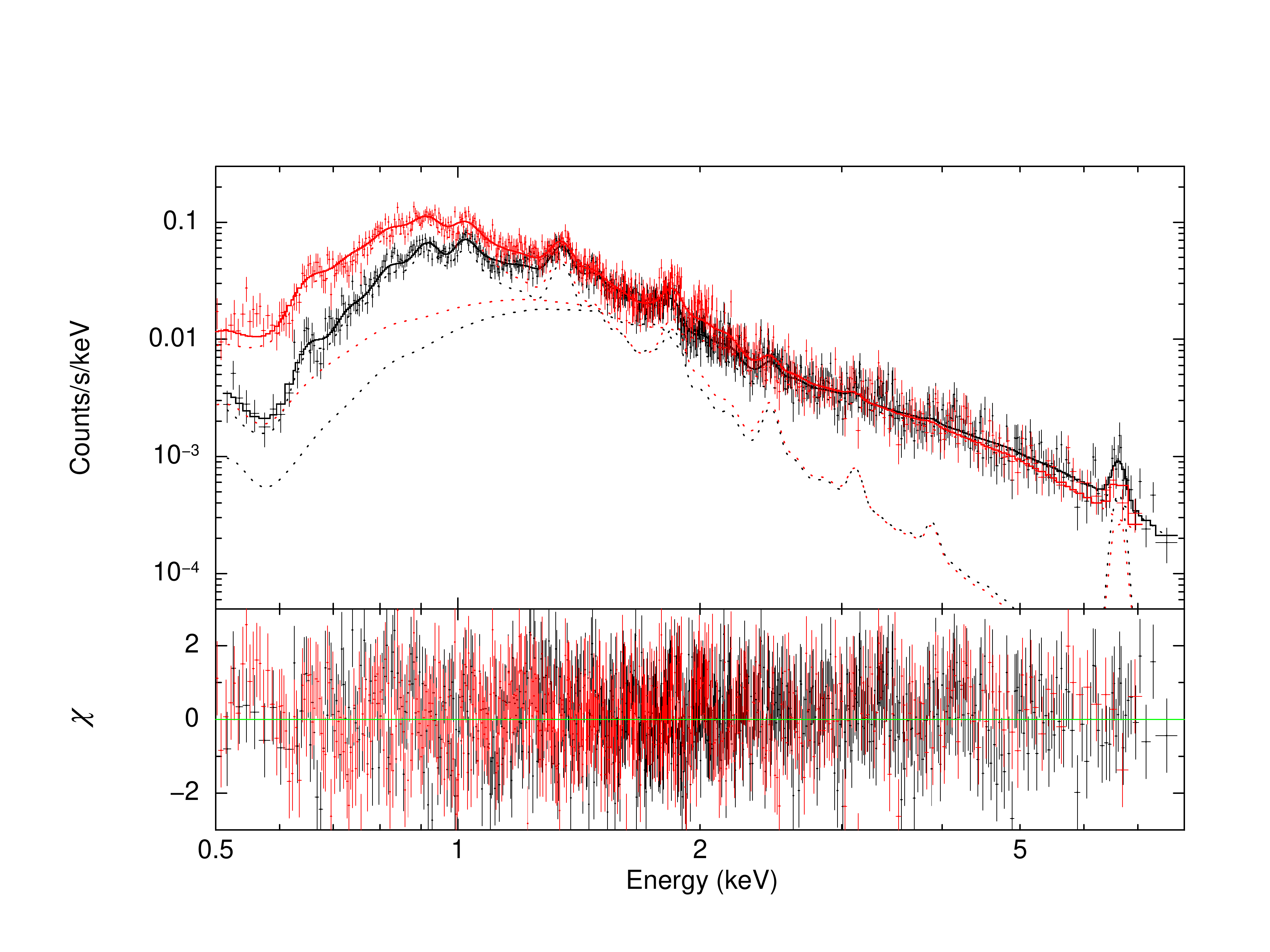}
\caption{\suzaku\  spectra of the central region and the best-fit model. 
The multiple components of the model are shown separately by the dotted lines:
seen at $\sim 1$~keV, for each type of CCD, the lower component is a power-law model and the upper one is a thermal plasma model with log-normal temperature distribution and log-normal foreground absorption. At $\sim 6.7$~keV, a Gaussian model is added to characterize the line emission.}
\vspace{0.3cm}
\label{f:spec_c}
\end{figure}

The above log-normal modeling for the temperature and absorption also reasonably well describes the spectra of the central region (Fig,~\ref{f:spec_c}; Table~\ref{t:spec-cen}). The fitted X-ray-absorbing gas column density and metal abundances are higher than those in the off-center regions. Although the mean plasma temperature is marginally lower than that in the off-center nebula region, the temperature dispersion is significantly larger. However, to fit the 6.7-keV Fe line, which the soft plasma component alone hardly accounts for, we have here included a Gaussian emission line in the modeling. In addition, the power law component appears substantially more prominent in the central region than in the off-center regions. The flux ratio of the power law to the plasma components is $\sim 0.48$ in the central region vs. $\sim 0.12$ in the off-center region. 

\begin{figure}
\includegraphics[width=1\linewidth,trim={0 30 10 100}]{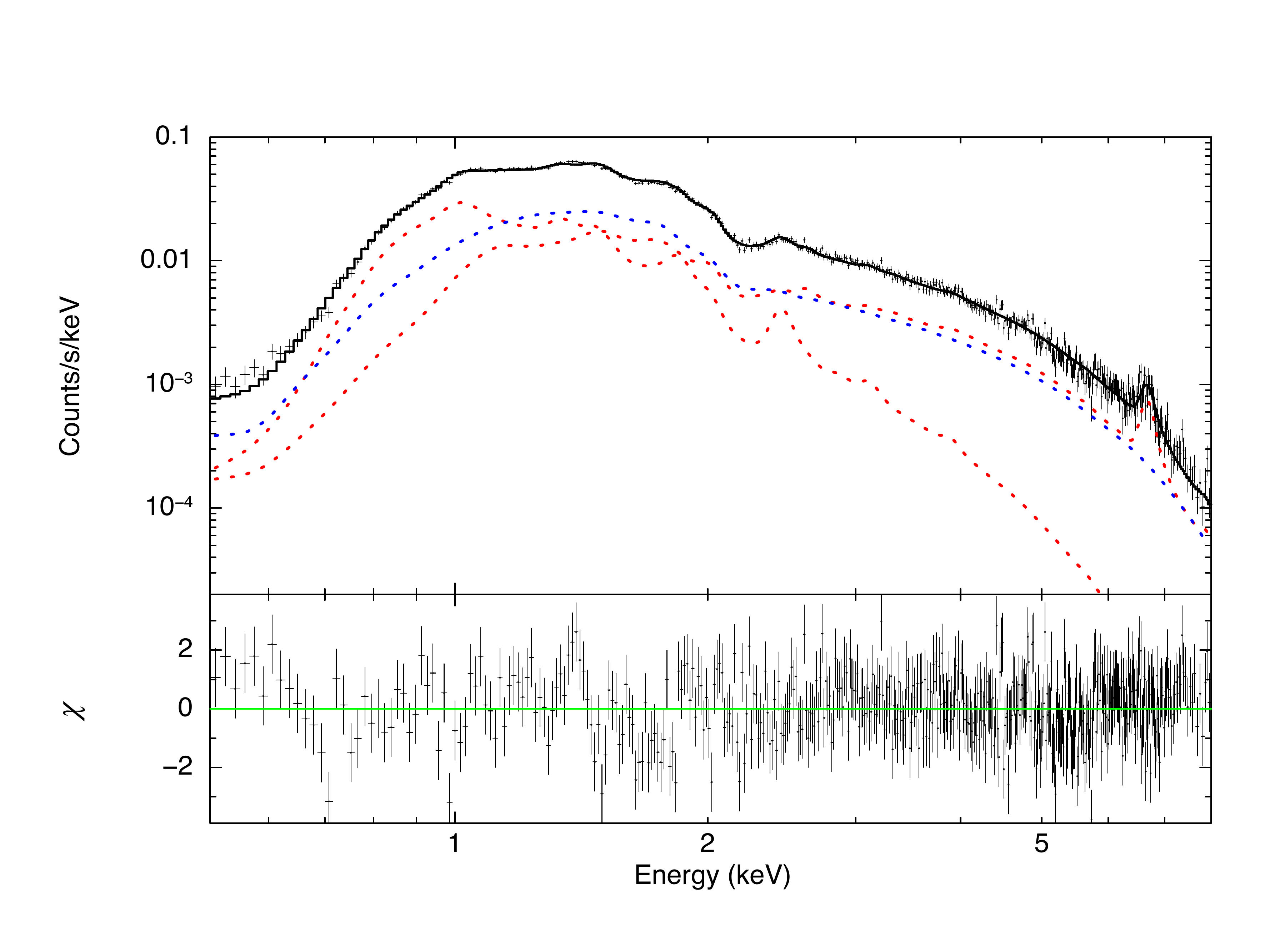}
\caption{Combined \chandra\ spectrum of detected sources within the central $1^{\prime}$ radius of NGC2070, together with the respective best-fit model (Table \ref{t:spec_chandra}). The two thermal plasma components with the log-normal temperature distributions and the power-law component are represented by the two red dashed lines and the blue dashed line, respectively.}
\label{f:spec_chandra}
\end{figure}

Apparently, the contribution from discrete X-ray sources in the central region is important. So we first check whether or not the enhanced power law component and/or the 6.7-keV Fe line emission can be explained by point-like sources detected in the \chandra\ data.  As can be seen in Fig.~\ref{f:spec_chandra} and Table~\ref{t:spec_chandra}, the \chandra\ source spectrum can be well characterized by a model consisting of two thermal plasma components plus a power-law component. Both the thermal plasma, as well as the foreground absorption, are assumed to have the log-normal distributions. The fitted photon index is $\sim 2.9$, which is quite comparable to $\sim 2.6$ obtained from fitting the \suzaku\ spectrum of the central region. 

\begin{figure}
\includegraphics[width=1\linewidth,trim={160 10 120 40}]{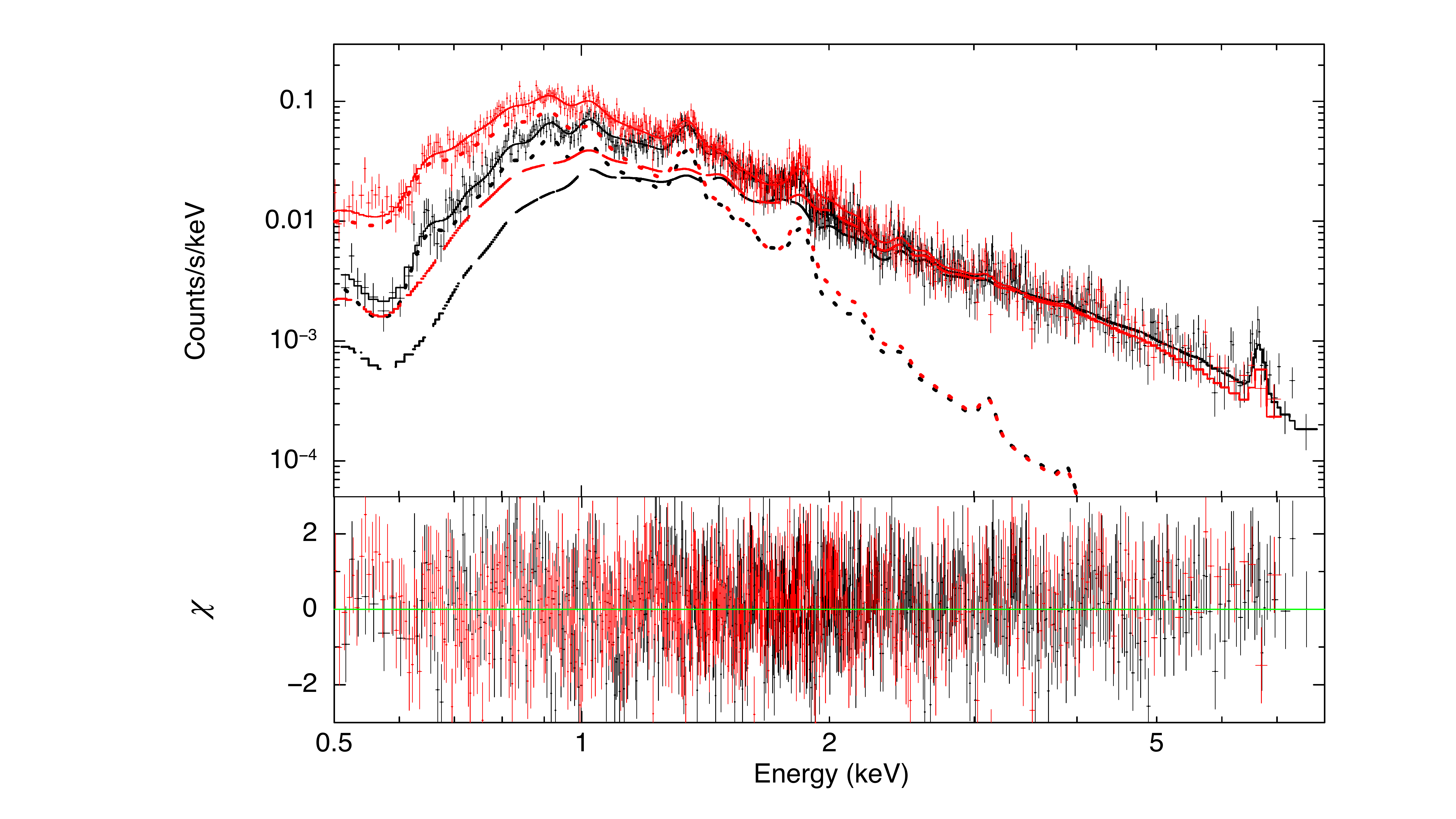}
\caption{The same \suzaku\ central region spectra as in Fig.~\ref{f:spec_c}, together with a best-fit model in which the power law component is replaced by a scaled-up best-fit \chandra\ point source model (Fig. \ref{f:spec_chandra}; Table \ref{t:spec_chandra} last column). The log-normal thermal components are marked by the dotted lines while the scale-up best-fit model of the discrete sources are represented by the dashed lines.}
\label{f:center_point}
\vspace{0.2in}
\end{figure}

To conduct a quantitative check of the discrete source contribution, we replace the power law and Gaussian components with the best-fit \chandra\ source model spectrum in the analysis of the\suzaku\ spectra. The model itself is fixed, but a multiplicative factor (the XSPEC constant model) is fitted in the analysis. The quality of the fit with this replacement  (Fig.~\ref{f:center_point}) is hardly changed, especially accounting for  the reduced number of the fitting parameters  (Fig,~\ref{f:spec_c}; Table~\ref{t:spec-cen}). However, this fitted factor is $9.7\pm0.3$, which is surprisingly much greater than the expected value ($\sim 1$). The factor should not be much affected by the area difference  (by a factor of $\sim4.6$)  between the regions used to extract the \suzaku\ and \chandra\ spectra, if the sources, which are very concentrated toward the center, are indeed the major contributor. We further test the effect due to the different  background subtraction methods, by changing the \suzaku\ off-nebula background to a local (on-nebula) background estimated in the annulus between $\sim2^{\prime}$ and $\sim9^{\prime}$ radii around NGC2070. This annulus contains little discrete source contribution, but has a high diffuse X-ray intensity, as seen in the \chandra\ image (Fig.~\ref{f:chandra}).  
With this local background subtraction, the factor decreases to $\sim 6$. This decrease suggests that the presence of a diffuse hard X-ray component, which is represented by the power law component in the above spectral analysis for the off-center nebula regions, can mimic a discrete source contribution. We may expect that this diffuse hard X-ray component is enhanced in the central nebula region and is thus not fully accounted for in our \suzaku\ spectral analysis for the central nebular region, even with the local background subtraction. So we conclude that a good fraction of the power law and Gaussian components in the central nebular region is diffuse in origin. The exact fraction is uncertain, however, because of the \chandra\ spectrum may not fully account for discrete sources in the central nebula region.

There are multiple arguments for the incomplete accounting of the discrete source contribution by the \chandra\ spectrum. One is that sources fainter than our source detection limit are not included in the spectrum. Even the detection limit itself is uncertain in such a crowded region as NGC~2070. The limit depends on the overall local background and source confusion, which varies with position and cannot be easily quantified. Another effect is the flux or luminosity variability of detected discrete sources. Several bright massive stars in the central region are known to show strong luminosity variability. As an example, we here consider Melnick 34 at R.A., Dec. (J2000) $= 05^h 38^m 44^s, -69^{\circ}06^{\prime}06^{\prime\prime}$, which is the most prominent discrete source in the central region. \citet{Pollock} show that this source most likely represents an eccentric Wolf-Rayet colliding-wind binary with a 155.1 day orbital period. The average luminosity of the binary is $1.2\times10^{35}~{\rm ergs~s^{-1}}$, although it can be a factor of up to 3 brighter or much fainter over a short interval of the orbital phase. The \chandra\ observations sample the cyclic light curve of the source quite well. Thus the accumulated \chandra\ spectrum should represent the binary with the average luminosity. Checking the orbital phase coverage of our \suzaku\ observation, we find that it should sample a luminosity of Mk 34 as about $1.5 \times10^{35}~{\rm ergs~s^{-1}}$, only somewhat higher than the averaged value. Thus this known variability cannot seem to explain the large luminosity difference between the \suzaku\ hard X-ray component and the accumulated \chandra\ source spectrum in the central region. More variability studies of the sources in the region is certainly needed.  But one probably cannot be sure about the exact contribution from the discrete sources, if they are not individually detected in any particular observation. The source detection limit in the Suzaku dataset varies across the field, but is down to about $2 \times 10^{34} {\rm~ergs~s^{-1}}$ in the nebula, except for the central region,  where the source confusion dominates, preventing us from excluding the possibility for an unresolved flaring/outbursting source to make an enhanced contribution to the power law component.  Therefore, for now, we conclude that the hard component (including both the power-law and the 6.7-keV line)  in the central region represents the combined contribution from a diffuse component and discrete sources, detected and undetected.

\subsection{Entire nebula}\label{s:entire}

\begin{figure}
\includegraphics[width=1\linewidth,trim={30 30 20 120}]{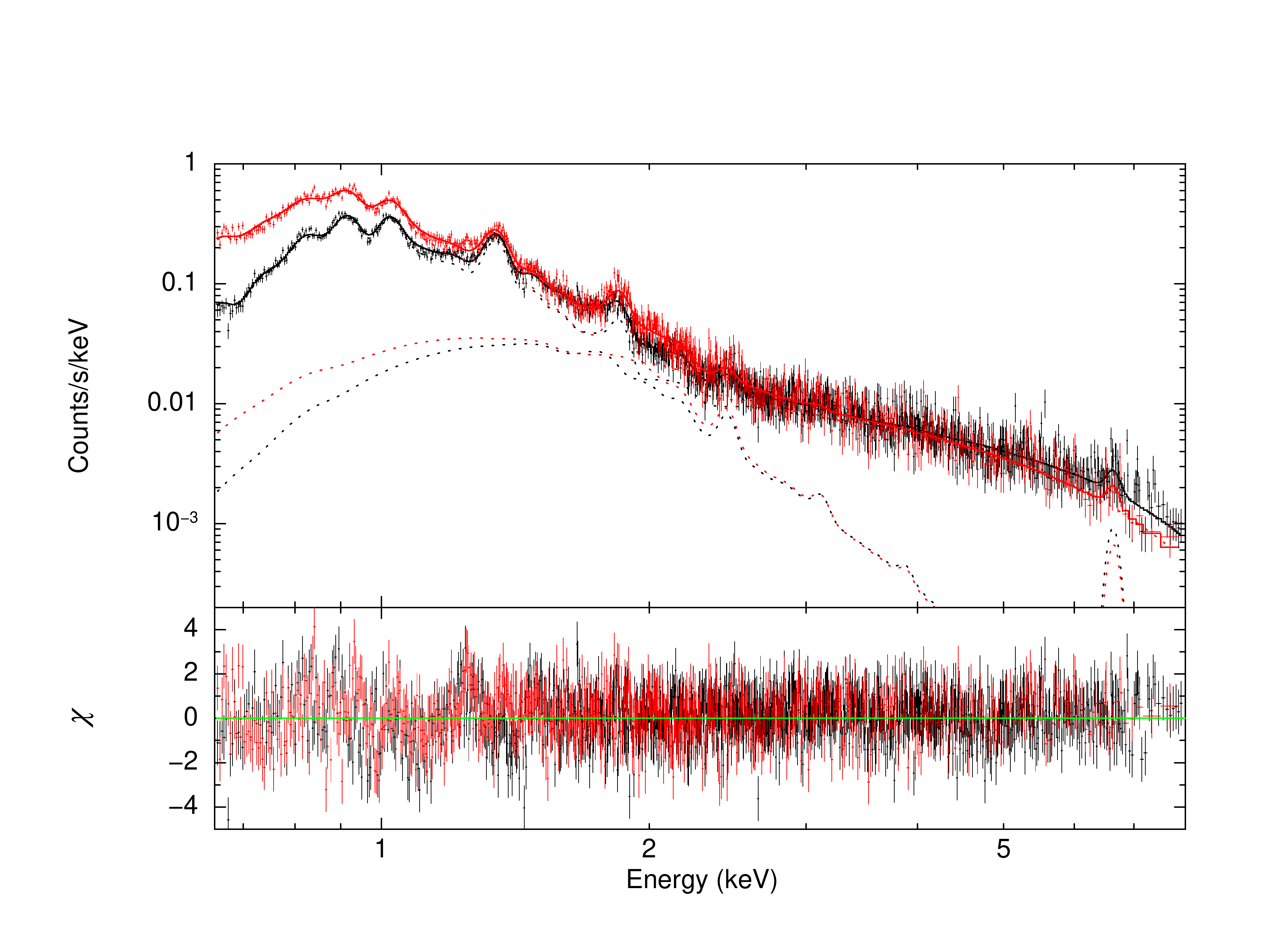}
\caption{\suzaku\ spectra of the entire nebula and the best-fit model in 0.65-8 keV, which is a log-normal thermal plasma with log-normal foreground absorption, plus a power-law component and a Gaussian line at $\sim 6.7$~keV. The rest is the same as in Fig,~\ref{f:spec_c}.}
\label{f:spec_entire}
\end{figure}

Interestingly, even the spectrum of the entire nebula can be quite well characterized by the log-normal temperature and absorption modeling. Table~\ref{t:spec-total} and Fig.~\ref{f:spec_entire} present the spectral fitting results.
The $\chi^2/dof$ value, though slightly greater than those for individual regions, is smaller that for the entire off-center nebula.  Naturally, the parameter values are in general between, or are statistically consistent with, those of the off-center and center regions.

\begin{table}
\caption{Spectral analysis results of the entire nebula}
\label{t:spec-total}
\begin{tabular}{lr}
\hline
Parameters & Values \\
\hline
$N_{\rm H, mean} (10^{22}\rm{cm}^{-2})$ & $0.41(0.39-0.43)$ \\
$\sigma_{N_{\rm H}} (10^{-2})$ & $1.5(1.0-2.2)$ \\
$\rm{EM(10^{59}cm^{-3})}$ & $57(52-69)$ \\
$kT_{\rm mean}(\rm{keV})$ & $0.31(0.28-0.32)$ \\
$\bar{x}\ $ &$1.28(1.18-1.31)$ \\
$\sigma_{x}\ $ & $0.49(0.45-0.52)$ \\
Ne (solar)& $0.9(0.8-1.1)$ \\
Mg (solar)& $0.7(0.6-0.9)$  \\ 
Al (solar)& $0.6(0.5-0.8)$  \\
Si (solar)& $0.4(0.3-0.5)$ \\
S (solar)& $0.5(0.4-0.6)$ \\
Fe (solar)& $0.3(0.2-0.4)$  \\
$\alpha$ (power-law index)& $1.91(1.86-1.96)$ \\
$\rm{Norm^{pow}}$ $(10^{-3}$$\rm{keV^{-1}}$ $\rm{cm^{-2}}$ $\rm{s^{-1}})$& $4.3(3.9-5.2)$  \\
$\rm{F^{pow}}$ $(10^{-12}$ ergs $\rm{cm^{-2}}$ $\rm{s^{-1})}$ & $6.61(6.52-6.70)$ \\
$\rm{F^{obs}}$ $(10^{-11}$ ergs $\rm{cm^{-2}}$ $\rm{s^{-1})}$ & $4.41(4.39-4.42)$ \\ 
$\chi^2/dof$ & $\frac{2085}{1788}=1.17$ \\
\hline
\end{tabular}

Note: The fitting is applied to the 0.65-8 keV band. A $\sim6.6$~keV Gaussian line is added to characterize the Fe line emission.
\end{table}

Finally we infer the physical parameters of the plasma in \xs, following the approach described in Appendix~\ref{a:para}. The dimensions of the individual regions and the entire nebula, as listed in Table~\ref{t:para_derived}, are used to estimate their volumes. For those elliptical regions, we consider the two approximations, prolate and oblate ellipsoids, as the lower and upper limits to the total volume $V_{T}$. Although part of the volume is likely to be occupied by cool gas, its filling factor should be small, compared to the overall uncertainty in the volume estimates. 

With $V_T$ and other measured parameters (Table~\ref{t:para_derived}), we infer the thermal energy as $E_{th} \sim (2.75-4.35) \times 10^{52} f_h^{1/2}
{\rm~ergs}$, where $f_{h}$ is the effective filling factor of the plasma in $V_{T}$. 

\begin{table*}
\caption{Several derived quantities for each region$^{\rm a}$}
\begin{tabular}{lccccccc}
\hline
 & Region1 & Region2 & Region3 & Region4 & Region5 & 1+2+3+4+5$^{\rm b}$ & Entire Nebula\\
\hline
$\rm{Dimension\ (pc)}$ & $31\times31$ & $31\times31$ & $31\times31$ & $50\times23$ & $42\times25$ && $104\times50 + 138\times36 $\\ 

[major $\times$ minor] & $ $ & $ $ & $ $ & $ $ &$ $\\
$\rm{Volume\ (10^4~pc^3)}$ & $12.5$ & $12.5$ & $12.5$ & $11.1/24.1$ & $11.0/18.5$ & $59.6/80.1$ & $145/363$\\
$\rm{Mass\ (M_{\odot})}$ & $3840$ & $2330$ & $3250$ & $2190/3230$ & $2130/2760$ & $13740/15410$ & $11620/18390$\\
${E_{th}\ \rm(10^{52}~ergs)}$ & $0.85$ & $0.62$ & $0.54$ & $0.45/0.67$ & $0.55/0.72$ & $3.01/3.40$ & $2.75/4.35$\\
${P_{th}\ \rm(keV~cm^{-3})}$ & $0.96$ & $0.70$ & $0.61$ & $0.58/0.39$ & $0.71/0.55$ & $0.71/0.64$ & $0.27/0.17$\\
${E_{th}^{single^{\rm c}}\ \rm(10^{52}~ergs)}$ & $0.72$ & $0.55$ & $0.48$ & $0.41/0.60$ & $0.50/0.65$ & $2.66/3.00$ & $2.15/3.40$\\
${P_{th}^{single}\ \rm(keV~cm^{-3})}$ & $0.86$ & $0.65$ & $0.57$ & $0.55/0.37$ & $0.68/0.52$ & $0.66/0.59$ & $0.22/0.14$\\
\hline
\end{tabular}

Note:  ${}^{\rm a}$ Two values, separated by a slash, are given for elliptical regions, corresponding to the assumed prolate or oblate ellipsoidal volume. ${}^{\rm b}$ in this column, $\rm{P_{th}}$ and $\rm{P_{th}^{single}}$ take the average value of the five individual regions, other items simply represent the values summed over individual regions.
${}^{\rm c}$ the superscript 'single' refers to the results from the single-temperature and single-column-density modeling.
\label{t:para_derived}
\end{table*}

\section{Discussion}
\label{s:dis}

Using the \suzaku\ observation, we have carefully examined the X-ray emission and absorption properties of \xs\ on various scales. Here we first compare our results with previous works and propose an X-ray spectral modeling approach for distant starburst regions. We then estimate the mass-loading and metal enrichment of the hot plasma, as well as the overall energetics of \xs. These estimates are confronted with the expected stellar feedback from NGC~2070. Finally, we explore the nature of the hard X-ray emission component and the evidence for significant cosmic ray (CR) acceleration in \xs.  

\subsection{Comparison with previous studies}
\label{ss:com}
Being the brightest HII region complex in the Local Group, \xs\ has been investigated in X-ray since 1990s.  The initial study based on data from the {\sl Einstein} X-ray Observatory revealed the presence of diffuse hot plasma enclosed in \xs~\citep{Wang2}. With very limited quality of the data, the emission spectrum was characterized with a simple isothermal plasma of a temperature $\sim 5 \times 10^6$ K and a total thermal energy of $\sim 3 \times 10^{52}$~ergs. The structure and evolution of the hot plasma were further investigated with {\sl ROSAT} observations 
\citep{Wang2}. \citet{Townsley1} presented a high-resolution \chandra\ observation of the nebula, showing region-to-region variations in the temperature range of $(3-9) \times 10^6$~K and in the X-ray-absorbing column density of $(1-6)\times10^{21} {\rm~cm^{-2}}$, and the absorption-corrected X-ray intensity of $(3-126)\times10^{31} {\rm~ergs~s^{-1}~pc^{-2}}$ in the 0.5-8~keV band. These basic results with large error bars are consistent with what we have obtained from the spectral analysis of the \suzaku\ data under the same assumed isothermal plasma and single foreground absorption model (e.g., Table~\ref{t:spec_indi}).

We have further used the log-normal distribution of the plasma emission measure as a function of temperature and/or the differential foreground absorbing column density to improve the X-ray spectral modeling. We have demonstrated that such a more realistic modeling, accounting for plasma temperature and X-ray absorption variations, is desirable or even necessary to more reliably infer the intrinsic properties (e.g., metal abundances) of a giant HII region such as \xs, especially when spatially resolved spectroscopy is not available. Moreover, if a plasma is significantly non-isothermal, a spectral fit assuming a 1-T plasma, for example, could lead to a substantial under-estimate of the emission measure of the plasma (see \S~\ref{ss:res-ind}), because the X-ray emission (proportional to the emission measure) may be dominated by relatively high-density and most likely low-temperature structures, which may account for a small fraction of the entire volume (Appendix~\ref{a:para}). The direction of the bias on the thermal pressure or energy, which depends on both the temperature and density, is demonstrated in Table \ref{t:para_derived}. 

The largest uncertainty in the estimation of the thermal energy or pressure remains to be due to different assumptions about the total volume and/or filling factor of the hot plasma.
When the entire nebula is concerned, the log-normal distribution modeling leads to relatively higher thermal energy and pressure estimates. while this estimate of the thermal energy is consistent with the sum of the corresponding values of the individual regions (well within their respective uncertainties),  the thermal pressure of the entire nebula is significantly smaller than any of the individual regions, owing to the sensitivity of the estimation to the filling factor as well as the volume size assumed ($P_{th}\propto (Vf_h)^{-\frac{1}{2}}$). For individual regions, which represent the brightest parts of  \xs\  to optimize the S/N ratios of the spectroscopy, their line-of-sight sizes, hence volumes, are likely underestimated, leading to over-estimates of their pressures. In contrast, the entire volume of the nebula tend to be overestimated; or equivalently, $f_h$  can be considerably smaller than 1. To bring the pressure estimate over the entire nebula to the value averaged over the individual regions, $f_h$  needs to be smaller than $\sim 0.2$.
In reality, of course, the pressure likely varies from one region to another. It is comforting to see that our thermal pressure estimates is consistent with the mean value measured in the central region of \xs\ via optical spectroscopy~\citep{Pellegrini11}. 

We propose that the log-normal temperature plasma plus long-normal absorption modeling represents an improved spectral characterization of the diffuse X-ray emission from a starburst region like \xs. The modeling is more realistic than the commonly used 1-T plasma plus a uniform foreground absorption model.

\begin{table*}
\caption{Metal abundance comparison : SB99 prediction v.s. spectral fitting}
\begin{tabular}{lcccc}
\hline
 & O & Mg & Si & Fe \\
\hline
${\rm SB99}$ & $0.57$ & $0.45$ & $0.36$ & $0.34$\\

${\rm Data^{\rm a}}$ & $0.4(0.3-0.6)\ $ & $0.7\ (0.6-0.9)$ & $0.4\ (0.3-0.5)$ & $0.3\ (0.2-0.4)$\\
\hline
\end{tabular}
\label{t:abundance}

Note:  ${}^{\rm a}$ the best-fit parameters for the entire nebula taken from Table \ref{t:spec-total} (the abundance of O cannot be constrained by the 0.65-8 keV entire nebula spectra, thus is taken from the fitting of total off-nebula spectra shown in Table \ref{t:off-center}), with $90\%$ confidence parameter ranges given in the parentheses. 
\end{table*}

\subsection{Estimation of the mass-loading from the ISM}

Mass-loading from the surrounding ISM strongly affects the emission and evolution of hot plasma in a superbubble. Theoretically, the mass-loading efficiency remains very uncertain and is related to a complex of processes, including turbulent mixing and thermal conduction,  as well as to the clumpiness of the ISM. Therefore, it is useful to make an empirical estimate of this efficiency, which can be obtained by comparing the mass of the hot plasma (inferred from X-ray spectroscopic analysis above) with the accumulated mass ejection from stellar winds and SNe in \xs. This latter mass ($M_{e}$) depends on the star formation history of \xs, which is somewhat uncertain, varying from one publication to another. \citet{Sabbi} recognize two distinct components of the young stellar population at the center of the nebula: R136 of $\sim 2$~Myr old and the so-called northeast (NE, hereafter) clump of $\sim 4$~Myr old. With these assumptions of the ages and the overall metal abundance to be $0.5$ solar for the ISM and hence young stars in the LMC \citep{Pellegrini11}, the running of the software Starburst99 v7.0.1 (SB99 hereafter, \citealt{Leitherer}) gives $M_{e} \sim (260+2324\times f_{\rm NE})M_{\rm s, 4}$, where $M_{\rm s, 4}$ is the total mass of young stars $(\rm{M}_{\rm NE}+\rm{M}_{\rm R136}$) in units of $10^4~M_{\odot}$ and $f_{\rm NE} \equiv \rm{M}_{\rm NE}/(\rm{M}_{\rm NE}+\rm{M}_{\rm R136})$ is the mass fraction of the NE clump.
These stellar masses remain somewhat uncertain, although a rough estimate suggests $(4-5)\times10^4~M_{\odot}$ and $(2-2.5)\times10^4~M_{\odot}$ for the R136 and NE components respectively (Elena Sabbi, private communications). Adopting $M_{\rm s, 4} \sim 7$ and $f_{\rm NE} \sim 0.3$, we obtain $M_{e} \sim 6700~M_{\odot}$ (of which $83\%$ arises from stellar winds). Comparing this $M_{e}$ value with the total mass estimate of the hot plasma in \xs, $M_{p}\sim (11620-18390)~M_{\odot}$ (Table \ref{t:para_derived}), we find the mass-loading fraction ranges from $(84-140\times f_{\rm NE})\% \sim 42\%$ to $(90-88\times f_{\rm NE})\% \sim 64\%$. 

To test the sensitivity of this result to the assumed stellar age distribution, we alternatively estimate $M_{e}$ , using a more continuous SFH of \xs, which is electronically readable and is made available by \citet{Schneider}. We numerically integrate $M_{e}$ over the age range of 0 to 6 Myr (the SFR decreased  rapidly beyond 6 Myr), and the resulting $M_{e}$ is 6380$~M_{\odot}$ for $M_{\rm s, 4} \sim 7$. The corresponding mass-loading fraction ranges from 45\% to 65\%. Therefore, we adopt the mass-loading fraction as $\sim 55\%$ with an uncertainty of $\sim 10\%$, which appears insensitive to the exact SFH assumed.

\subsection{Comparison with stellar metal enrichment}

\begin{figure}
\includegraphics[width=1\linewidth]{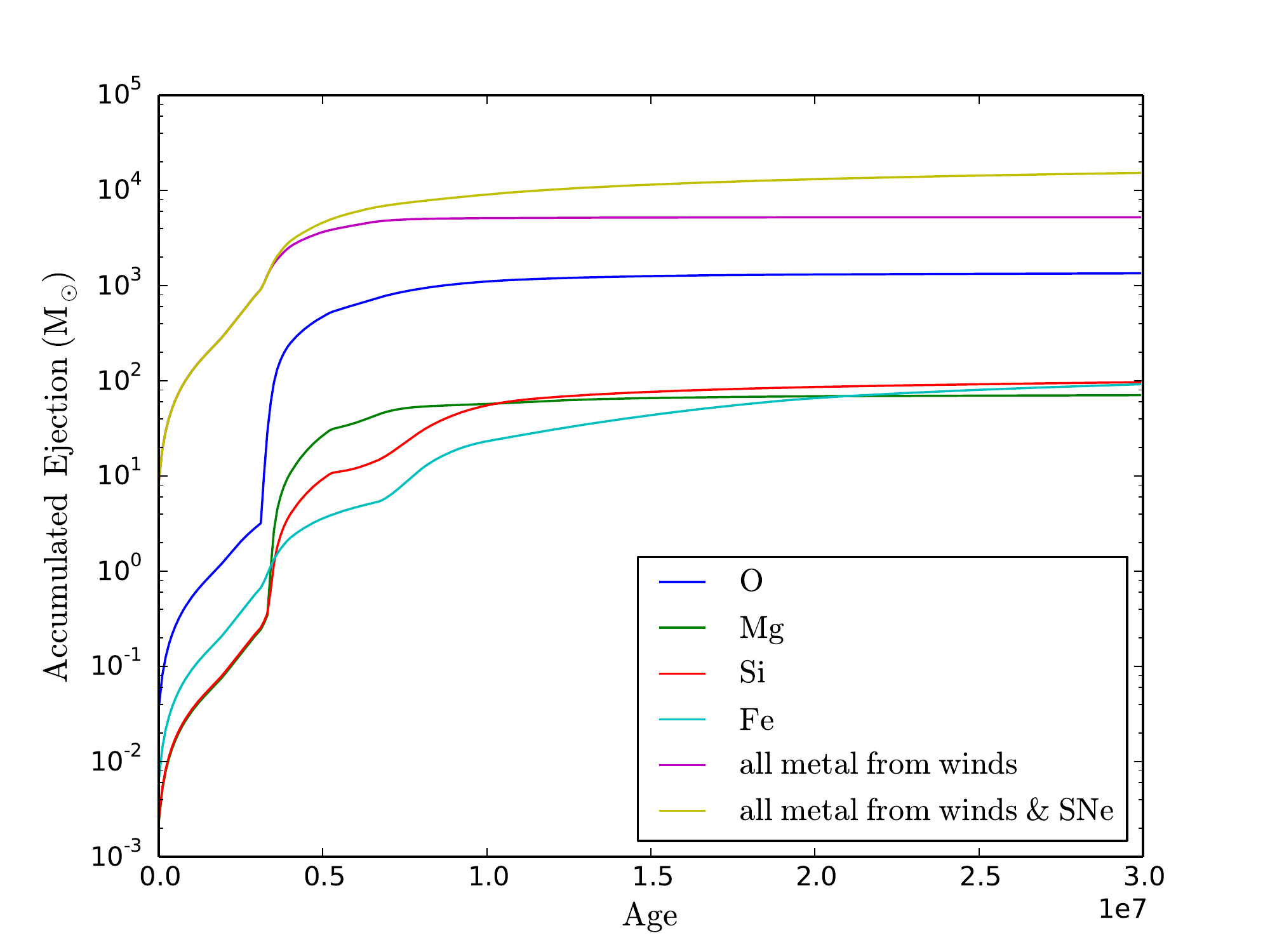}
\caption{Accumulated metal ejection from a star cluster with a total stellar mass of $1\times10^5M_{\odot}$ as a function of age.
}
\label{f:elem}
\end{figure}

\begin{figure}
\includegraphics[width=1\linewidth]{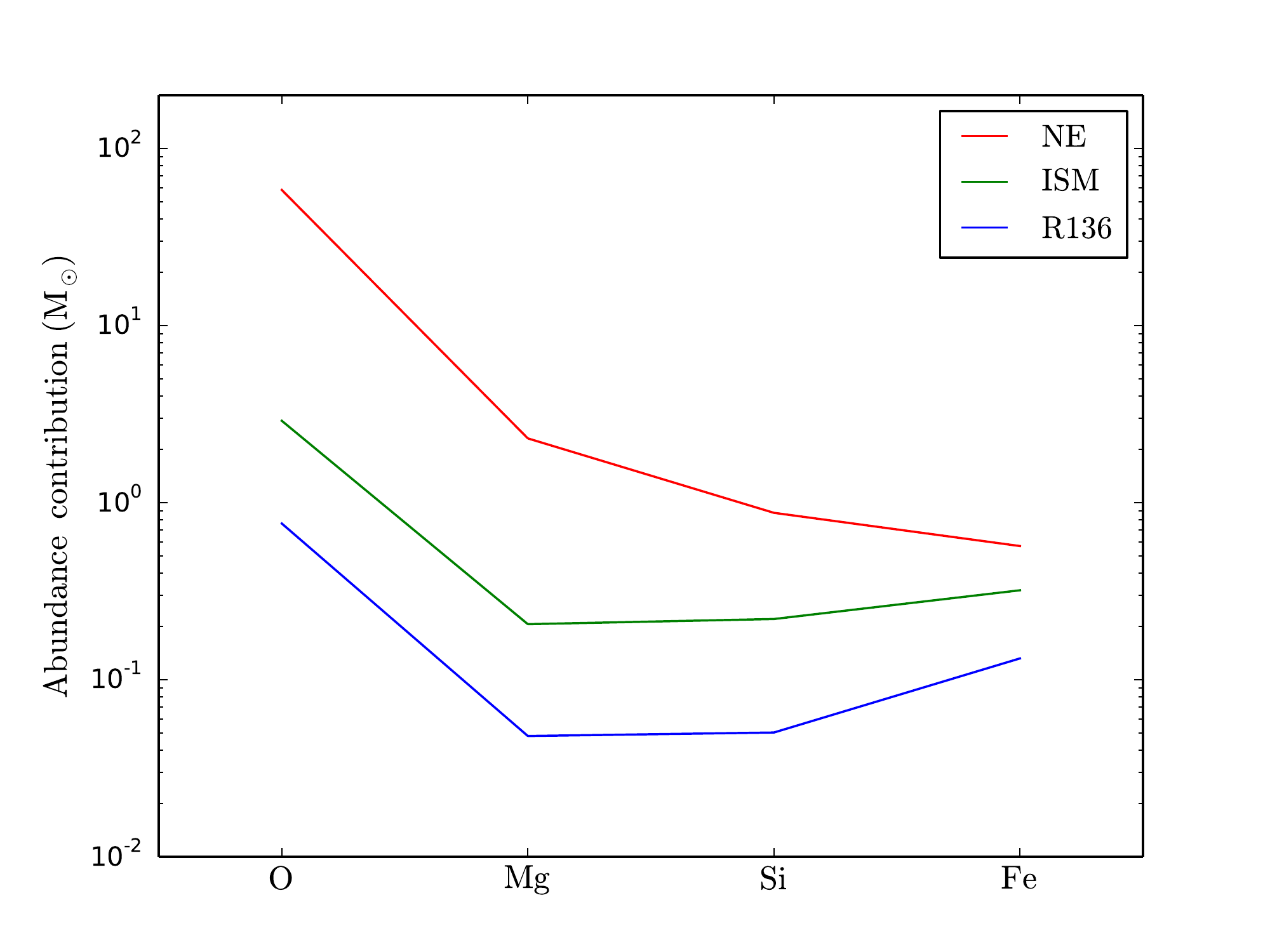}
\caption{Total expected metal abundance contributions from the surrounding ISM (green), as well as the NE cluster (red) and R136 (blue) at their present ages. Four key elements (O, Mg, Si, and Fe) which have been estimated in our spectral fits are included.}
\label{f:abun}
\end{figure}

With the mass-loading fraction, we can further check the consistency of our measured metal abundances with the chemical yield of the young stars in \xs. Again using SB99, we illustrate  in Fig.~\ref{f:elem} the accumulated metal ejection as a function of time, including a few key elements (O, Mg, Si, and Fe), separately. 
Fig.~\ref{f:abun} compares the elemental contributions from the NE and R136 stellar components and from the above adopted ISM mass-loading, $M_{X, {\rm SB99}}$ (per $10^5M_{\odot}$), where $X$ represents one of the elements considered here: O, Mg, Si, and Fe. The abundance of an individual element can be defined as 
\begin{displaymath}
\frac{(X/\rm{H})_p}{(X/\rm{H})_\odot} = \frac{\text{mass-loading fraction} \times M_{p}\times0.5(\frac{M_{X}}{M_{\rm H}})_\odot+0.7M_{X, {\rm SB99}}}{M_{p}(\frac{M_{X}}{M_{\rm H}})_\odot}
\end{displaymath} 
, where $M_p$ is the total plasma mass, and a metal abundance of 0.5 solar is adopted for the ISM in \xs\ \citep{Pellegrini11}. The derived abundances are demonstrated in Table~\ref{t:abundance}, with a comparison with our spectral fitting in Table~\ref{t:spec-total}. The derived abundances of O, Si and Fe are consistent with our spectral fitting results within the 90\% confidences, although the abundance of Mg is slightly smaller than the uncertainty range.

\subsection{Comparison with stellar energy feedback}

The coupling of  stellar energy feedback with its surrounding interstellar medium remains a key uncertain aspect in galaxy formation and evolution theories. In addition to the mechanical energy injection in form of stellar winds and supernovae, radiative transfer feedback from massive stars via direct or dust-processed radiation pressure and photo-ionization can also play a significant role in dispersing dense dusty gas and possibly in driving outflows from starburst regions. Studying the energetics of \xs\ allows us to test the efficiencies of these two forms of stellar feedback.

Theoretically, the effect of the mechanical energy injection on the surrounding interstellar medium is modeled as a superbubble. The combined energy input from fast stellar winds and supernovae drives the expansion of low-density hot plasma, which sweeps much of the surrounding gas into a dense cool shell. In a simple spherical superbubble model proposed by \citet{Mac}, 27/77 of the total energy input is lost in the formation of the cool supershell around the superbubble. The remaining 50/77 is split between the thermal energy of the plasma inside the superbubble (35/77) and the kinetic energy of the supershell (15/77). 

Based on the SB99 again, we estimate the total mechanical energy injected by NGC~2070. Adopting its two-age component characterization of 2 and 4 Myr and the ISM metal abundance of 0.5 solar, we obtain the total energy input as $\sim 8.3 \times 10^{52}$~ergs, of which only $\sim 1.3 \times 10^{52}$~ergs is from SNe. According to the superbubble model, we then expect a kinetic energy of $1.6 \times 10^{52}$~ergs of \xs. In the following, we estimate significant contributions from individual ISM components in HI, HII, and molecular phases. The results are summarized in Table~\ref{t:energy}.

While \citet{Chu} has presented a detailed study of the kinematics of HII gas and gives an estimate of its kinetic energy as $\sim 1.0 \times 10^{52}$~ergs, the estimation of the other two contributions needs some effort. We estimate the HI contribution, using the hydrogen 21 cm line emission survey of the LMC, made with the Australia Telescope Compact Array\footnote{$http://www.atnf.csiro.au/research/lmc\_h1/index.html$}. 
We first conduct a 2-D Gaussian fit to the HI emission data, which gives an intensity-weighted centroid in the sky, as well as the corresponding Gaussian widths, $\sigma_x$ and $\sigma_y$. We then use FWHM to define the on-nebula region as an ellipse with its major and minor semi-axis to be $\sqrt{2 \ln{2}}\sigma_x$ and $\sqrt{2 \ln{2}}\sigma_y$. The off-nebula region is defined as an elliptical annulus with its inner and outer radii being 2 and 3 times the FWHM; all are centered at the centroid. Following \citet{Kim}, we use the off-nebula region to estimate the local HI background, which can then be subtracted from the on-nebula region. The net mass of the HI gas associated with \xs\ is calculated as $M_{\mathrm{HI}}=1.8 \times 10^{18}~W~A~m_{\mathrm{H}} \approx 1.6 \times 10^{6}~M_{\odot}$, where $W$ is the average intensity in ${\rm K~km~s^{-1}}$, $A$ is the 2-D projected area of the nebula (the ellipse defined above) in cm$^2$, and $m_{\mathrm{H}}$ is the hydrogen atom mass. Similarly, we estimate the net intensity-weighted line-of-sight velocity dispersion of the HI gas as $\sigma_v\sim7.8~{\rm~km~s^{-1}}$. Thus, the kinetic energy of HI gas (assuming statistically isotropic motion) can be estimated as $E_K(HI) = \frac{3}{2}M_{HI}\sigma_v^2 = 2.9 \times 10^{51}$~ergs. 

We further estimate the kinetic energy in the molecular phase of the ISM associated with \xs. To do so, we use dust to trace the cool gas in both molecular and HI phases. \citet{Gordon} derived a map of dust mass  across \xs\ from the SED fitting over the far-IR to sub-mm range. We multiply this map by the constant gas-to-dust ratio of 340 ($\pm$ 40), which is an average value for the LMC \citep{Skibba}, to infer a total cool gas mass distribution. By subtracting from this distribution the HI mass map (after being convolved with a Gaussian kernel to match the spatial resolution), 
we obtain a molecular gas mass map. This map is arguably more reliable than one that could be inferred from the CO emission, as mapped out by the Magellanic Mopra Assessment \citep{Fukui}. We do use the CO emission to trace the line-of-sight velocity dispersion of the molecular gas in a way similar to that used for the HI component. The kinetic energy of the molecular gas associated with \xs\ is estimated as $E_K(H_2) \sim 5 \times 10^{50}$~ergs. In any case, the contributions of the HI and molecular gases to the total kinetic energy of the ISM in \xs\ seem to be much smaller than that of the HII gas (Table~\ref{t:energy}). 

\begin{table}
\caption{Estimated energetics for various gas phases}
\begin{tabular}{lc}
\hline \hline 
Phase & Energy $\left(10^{52} \mathrm{ergs}\right)$ \\
\hline 
HII & 1.0 \\
HI & 0.3 \\
Molecular & 0.05 \\
Total Kinetic & 1.35 \\
\hline 
Thermal & $\lesssim 3$ \\
\hline
Total & 4.35 \\
\hline
\end{tabular}
\label{t:energy}

\end{table}

From the above estimates of the kinetic energies in HII, HI and molecular gases in \xs, we find their sum as $\sim 1.35\times 10^{52}$~ergs (Table~\ref{t:energy}). This sum is comparable to the total kinetic energy ($1.6 \times 10^{52}$~ergs) predicted by the superbubble model (see above). Therefore, there is no evidence for a significant energy input from the radiative transfer feedback, which may otherwise substantially affect the motion of the cool gas. Nevertheless, the radiation pressure could still play a subtle role in the evolution of a giant molecular cloud (e.g., \citealt{Elmegreen, Scoville}). The contribution of the radiative transfer feedback in reversing the collapse of a proto nebula is typically not considered in the modeling of superbubble evolution. While the radiation pressure is inescapable during the formation phase of a stellar cluster, the resulting outward expansion of the gas is at most a few ${\rm km~s^{-1}}$, and should decrease further after sharing the momentum with the swept-up materials and overcoming the gravity of the cloud. Therefore, in a starburst region similar to \xs, the ultimate contribution of the radiative transfer feedback to the total kinetic energy is probably insignificant, compared to the later mechanical energy inputs primarily from stellar winds and SNe.

Interestingly, the thermal energy, $3.8 \times 10^{52}$~ergs, predicted by the superbubble model may be  considerably larger than $E_{th} \sim (2.75-4.35) \times 10^{52} f_h^{1/2}{\rm~ergs}$ inferred from our spectral analysis (\S~\ref{s:entire}), because $f_{h}$ is likely to be much smaller than 1 (see \S~\ref{ss:com}). Therefore, we may conclude that there may be important energy loss channels that are not included in the simple superbubble model compared here.

\subsection{Plausible energy loss channels of the hot plasma}

 We here consider a few plausible channels of the energy loss. First,  the radiative cooling of the plasma, which is not significant at present, could be important in an early evolutionary stage of \xs. However, as long as the duration of this stage is short, the effect on the accumulated energy in the nebula is unlikely to be significant, because the mechanical energy input from massive stars is a rather continuous process, as is expected in the superbubble scenario \citep{Mac}. 

Second, significant heat loss may be due to the presence of dust in the hot plasma.  Indeed, both {\sl WISE} 22 $\mu$m and {\sl Spitzer}/MIPS 24 $\mu$m images  \citep[e.g.,][]{Meixner2006} show enhanced emission within \xs, preferentially at inner edges of H$\alpha$-shells, consistent with what is seen in Galactic HII regions  \citep[e.g,][]{Churchwell2009,Paladini2012}. The sputtering of dust grains takes time. Dust can be mass-loaded into the plasma from the surrounding cool ISM and can also form in supernova ejecta and in stellar winds, especially in colliding wind binaries. Some ejecta clumps may eventually escape into the surrounding medium before being sputtered \citep{Martinez-Gonzalez2018,Slavin2020}.  Hot electrons can lose energies via thermal collision with dust grains, which then radiate in mid-infrared.  The dust-depletion of certain elements  or their being carried away by dust could, in principle, lead to an underestimate of the metal abundances of the hot plasma.  But because of the relative large age and volume of \xs, compared with individual SNRs, these effects should be quite minor, although their quantification will need dedicated studies \citep[e.g.][]{Martinez-Gonzalez2017}.

Third, one may speculate that there could be some leakage of hot plasma from \xs\ into very low density regions (e.g.,  into the halo of the LMC). 
In this context, it is interesting to discuss the potential connection to the galactic wind from the LMC, as recently revealed by \citep{Ciampa2021} from Wisconsin H-alpha Mapper observations. The high-velocity component of this wind seems to spatially align with \xs, but on scale of $\sim 2^\circ$, which may be limited by the angular resolution ($\sim 1^\circ$ beam size) of the observations. We suspect that the wind seen in the field of \xs\ is largely driven by adjacent older and larger starforming regions, primarily LMC2 and LMC3 — two known blowout supergiant bubbles \citep{Wang1991,Points2001}. Their angular scales ($\sim 1^\circ$) and ages ($\sim 10^7$ yrs)  are reasonably consistent with the line-of-sight physical and outflow time scales inferred from the modeling of the wind. \xs\ itself is probably far too compact and young to explain these scales. We are also not aware of any evidence for a systematic nebula-wide flow of H$\alpha$-emitting gas from \xs\ at a velocity of $\gtrsim 100 {\rm~km~s^{-1}}$, as reported for the galactic wind. 

Finally, the superbubble model does not account for the possibility that a significant fraction (e.g., 10\%) of the input energy may be used to accelerate CR \citep{Butt}. 

\subsection{Evidence for cosmic ray acceleration in \xs?}
\label{ss:ntx}

A significant energy release via CR is supported by our tentative detection of a significant diffuse hard X-ray  emission from \xs, other than N157B. The presence of this emission is most convincing in the entire off-center region, where the confusion with discrete sources is minimal and the counting statistics of the \suzaku\ spectral data is still sufficiently high. The spectral shape of the emission is well characterized by a power law with a photon index $\lesssim 2$. In principle, this hard component may be mimicked by a plasma with an extreme high temperature ($\gtrsim 10$~keV) and an unusually low metal (mostly iron) abundance, which is considered to be far from being physical. Therefore, we conclude that the component most probably represents a diffuse NTX component in \xs. 

So far, the only convincing case for significant NTX emission from a superbubble is 30 Dor C with strong observational evidence over the energy range from radio to TeV~\citep{Bamba, Smith2004, HESS2015, Babazaki, Lopez2020}. This superbubble has a diameter of $\sim 100$~pc, much smaller than that of \xs. Also the NTX emission is confined to the western rim of the superbubble and is suggested to be due to a recent 
off-center SNR 
in the region (e.g., \citealt{Chu1990,Wang1991b}). The most plausible physical mechanism for the NTX  emission is synchrotron radiation of $\gtrsim 10$~TeV electrons \citep{Smith2004, Bykov}, which may also be responsible for the TeV emission via the inverse Compton scattering of ambient radiation field. The observed X-ray and TeV emissions from N157B can similarly be explained~\citep{HESS2015}.

In contrast, there is no evidence for a significant TeV emission and hence the presence of diffuse X-ray synchrotron-emitting leptons in \xs~\citep{HESS2015}. This may not be surprising. The cooling timescale of X-ray synchrotron-emitting electrons is only on order of years. If synchrotron radiation were the dominant process for the NTX emission, as has been proposed for both N157B and 30 Dor C, then it would directly trace where the acceleration is ongoing. There is probably no significant large-scale strong shock required to accelerate electrons to X-ray synchrotron-emitting energies efficiently (compared to the synchrotron-cooling) in \xs. 

The NTX emission from \xs\ could instead be due to inverse-Compton scattering of the strong infrared radiation field, for example. The relevant CR leptons have a characteristic Lorentz factor $\gamma \sim 10^2$, which should be easy to produce and have a long cooling timescale. A strong presence of the low-energy CR leptons in \xs\ may be expected, because generally SNRs in low-density hot gas within a superbubble should have relatively weak shocks~\citep{Tang2005, Sharma}.  

In comparison, protons and heavier ions can be effectively accelerated to high energies by repeated stochastic shocks in a superbubble (e.g. \citealt{Lingenfelter2019}). Both the size and timescale of \xs\ are substantially greater than those of individual SNRs, which favors the acceleration of ions to greater energies. Therefore, \xs\ can  be a significant site for CR production. 

\section{Summary}
\label{s:sum}
Based chiefly on the XIS data from our 100~ks \suzaku\ observation, we have conducted a systematic spatially resolved and integrated spectral analysis of \xs. 
We start with individual regions, which may be considered to have relatively uniform X-ray-emitting and absorbing properties, and then move on to large complexes, which need sophisticated modeling. The consistencies among different ways of spectral modeling and different scales have been checked. The implications of our results on various issues are discussed in \S~4. Our main results and conclusions are as follows:

\begin{itemize}

\item Both the thermal plasma temperature and the foreground X-ray absorption show significant variation across \xs. Although the spectra extracted from individual off-center regions (as defined in Fig.~\ref{f:im_reg}) can be well characterized by a thermal plasma of temperature in the range from $\sim$ 0.2-0.4~keV with simple foreground absorption, both the temperature  and the foreground absorption do vary significantly from one region to another. These variations can be reasonably well characterized by the log-normal temperature and absorption distribution models. This log-normal modeling is also desirable for the central and combined off-nebula regions to produce consistent results between spatially resolved and integrated analyses. Therefore, we propose that the log-normal modeling is suitable for the analysis of other giant HII regions, especially when a spatially resolved analysis is not feasible. 

\item We find that a substantial mass-loading from the ISM to the hot plasma is needed to account for $\sim 55\%$ of the measured plasma mass $\sim 1.5 \times 10^4~{\rm~M_{\odot}}$, which may be compared with $\sim 19\%$ of the stellar mass of the central OB association NGC~2070. This mass-loading is also required to explain the metal abundances of the plasma. The total contents of key metal elements in \xs\ are consistent with the expected chemical enrichment of NGC~2070, especially due to the presence of its NE clump ($\sim4$~Myr), which has contributed about 54 SNe according to the SB99 calculation.

\item The sum of the kinetic and thermal energies observed in the nebula (Table~\ref{t:energy}) is considerably smaller than the total mechanical energy expected from the feedback of NGC~2070. Interestingly, the total kinetic energy estimated from the HI, HII, and molecular gases in the nebula appears to be consistent with the feedback, according to a simple superbubble model. The observed shortfall appears mostly in the total thermal energy of the plasma, which we estimate to be $\sim 2.8-4.4 \times10^{52}f_h^{1/2}$~ergs, where $f_h$ is the filling factor and should be as small as $\sim 0.2$. The estimated relevant thermal pressure of the hot plasma is consistent with the independent measurement of HII gas in the central region of the nebula. Therefore, we conclude that there are likely important energy loss channels that the superbubble model does not account for. Such channels include the interaction of hot plasma with dust grains and the generation of CR.

\item We show evidence for a diffuse NTX component in \xs. Our \suzaku\ spectral decomposition indicates a power-law component in both the center and off-center regions. This component is considerably harder than the thermal plasma emission and can only be partly explained by discrete sources even in the central region, including the known variability of the brightest colliding-wind binary MK 34. The remaining contribution to the power law component likely arises from the diffuse NTX emission, probably representing inverse-Compton scattering of strong (reprocessed) stellar light radiation in the nebula. The exact amount of this contribution is somewhat uncertain, however, owing to the uncertainties in both the spectral decomposition of the X-ray emission and the contamination estimation from discrete sources. The NTX emission throughout much of the nebula, if confirmed, may suggest that CR is effectively accelerated in starburst regions which is not due to individual SNRs. 

\end{itemize}

\section*{Acknowledgments}

We thank the referee for constructive comments on the work, which was partially supported by the Space Telescope Science Institute (STScI/NASA) grant, HST-GO-15164.002-A.

\section*{DATA  AVAILABILITY}

The \chandra\ and \suzaku\  data as described in \S~\ref{s:obs} are available in the data archives accessible via https://heasarc.gsfc.nasa.gov/cgi-bin/W3Browse/w3browse.pl. Processed data products underlying this article will be shared on reasonable request to the authors.

\appendix
\section{Estimation of the X-ray background spectral contribution}
\label{a:spec_back}
To study the spectral properties of \xs, we need to estimate both X-ray and non-X-ray background contributions to each of the observed spectra. The X-ray background, assumed to have the same intrinsic spectral shape and surface intensity across the detector field of view, is subject to non-uniform instrumental effect, which is accounted for by the arf and rmf files of the spectrum. We characterize the X-ray background with the XSPEC model combination of 
$apec_1 + wabs(apec_2+powerlaw)$, where the first part \textit{$apec_1$} is an optically thin thermal plasma~\citep{Smith} to mainly account for the Local Bubble emission, while the second part \textit{$apec_2$}  is for the Galactic gaseous halo  and the collective contribution of unresolved sources (mostly AGNs), which is represented by a power law and is important mostly at energies $\gtrsim 1.5$~keV. \textit{wabs} is the foreground absorption with solar metal abundances assumed~\citep{Morrison}. The exact meaning of these components is not important here. More important is is how well the combination of them describes the overall spectral shape. 

\begin{figure}
\includegraphics[width=1\linewidth,trim={30 100 0 100}]{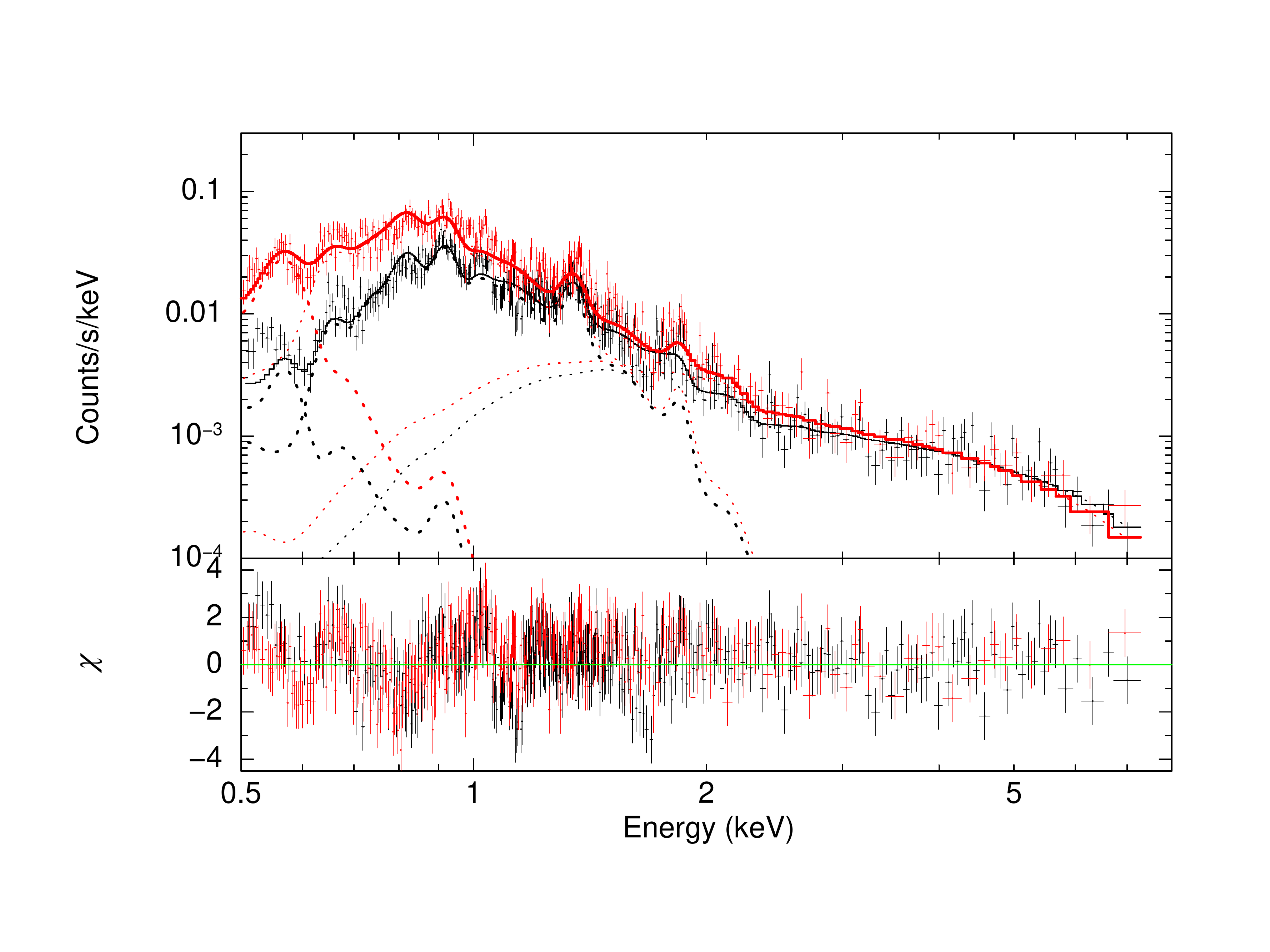}
\caption{The background spectra from the FI (black) and BI (red) detectors and the corresponding model fits (solid lines). The dotted lines show two-temperature thermal components (APEC) and power-law components separately. }
\label{f:spec-back}
\end{figure}

\begin{figure}
\includegraphics[width=3.6in,trim={30 30 50 80}]{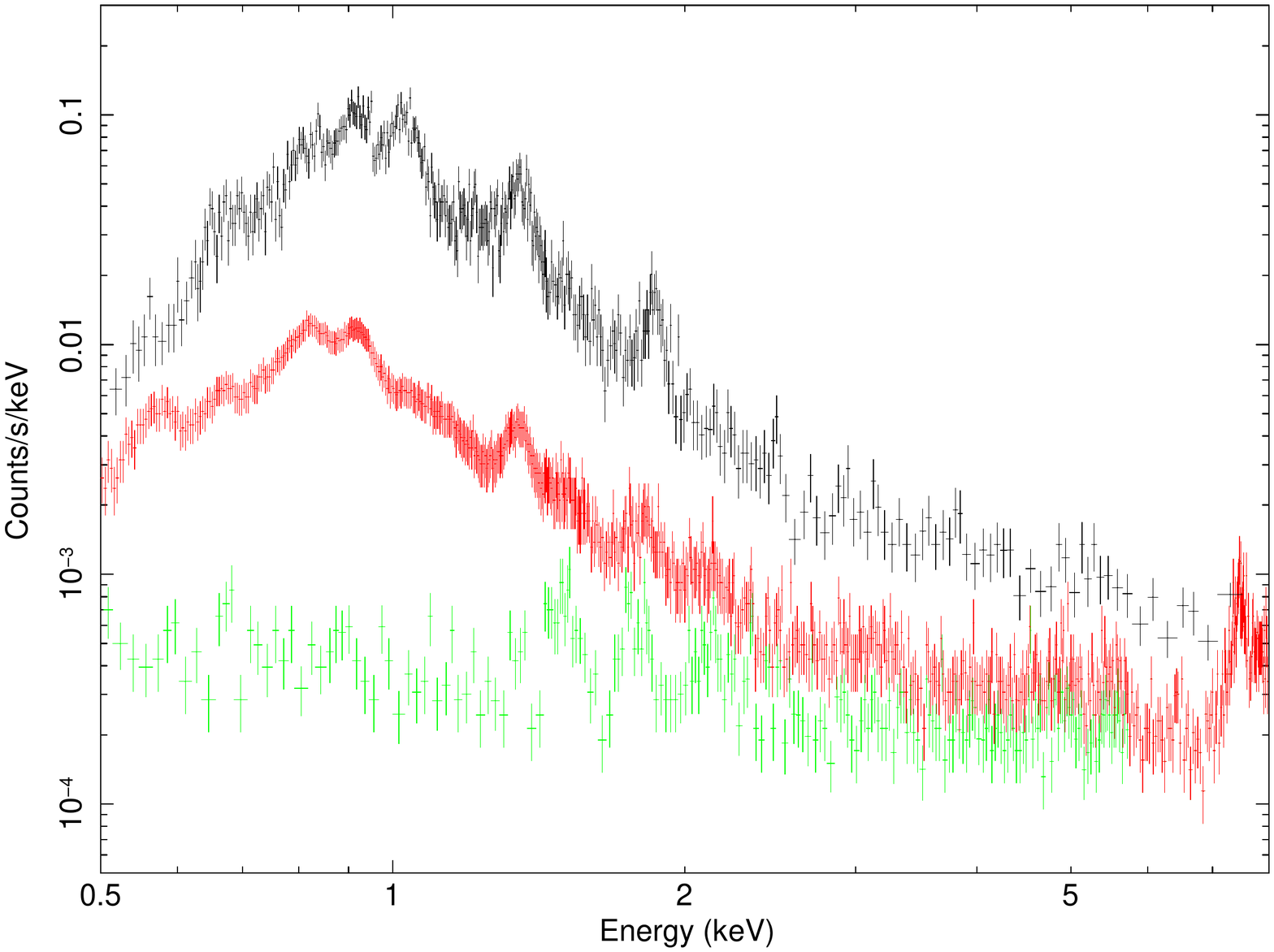}
\caption{Comparison of the XIS1 spectra in Region 2: the total spectrum without any background subtraction (black); the combined background spectrum (area-scaled local X-ray background + NXB (red); and the NXB (green). All three spectra are re-binned to achieve a minimum S/N ratio of 3.0.}
\label{f:compare}
\end{figure}

While Table~\ref{t:spec_back} lists the fitting results, Fig.~\ref{f:spec-back} compares the best-fit model to the off-nebula X-ray background spectrum. The relative large $\chi^2/dof\ =\ 1.37\ (941/689)$ is apparently due to some systematic differences between the data and the model (e.g., at the level of up to $|\chi| \sim 2$, e.g., the lower pannel of Fig.~\ref{f:spec-back}). A close look of the fit in Fig.~\ref{f:spec-back} also shows that there seems to be a systematic difference (up to a factor of $\sim 2$) between the FI and BI spectra in the 0.5-0.65~keV range, probably representing the uncertainties in the non-X-ray background subtraction at the lower energy end of the spectral coverage. To check how such uncertainties may affect  the background subtraction of on-nebula spectra, we compare the spectra of the total detected events with those of the estimated NXB and NXB+local X-ray background contributions (e.g., for Region 2 in Fig.~\ref{f:compare}). We find that these uncertainties can cause significant effects in the modeling of the on-nebula X-ray spectra, mostly at the low ($\lesssim 0.6$~keV) or  high ($\gtrsim 7$~keV) end of the energy coverage.

In the subsequent spectral analysis of on-nebula regions, we fix the X-ray background to its best-fit model scaled by the area ratio of a specific region to the total selected background field (as in Fig.~\ref{f:compare}). Alternatively, one may jointly fit the background with both the on- and off-nebula spectra, where the latter includes additional model components of varied sophistication (see the main text). The model spectral parameters for the background component do change (but not significantly in general) from one fit to another (for different regions and/or different on-nebula models).
Therefore, we adopt the simpler, more self-consistent and reproducible approach of fixing the background spectral shape to its best-fit model. 

With this best-fit X-ray background model, we estimate the X-ray background contribution to each of the spectra extracted from on-nebula regions. Specifically,  in XSPEC we 1) multiple the background model by a constant which is the on- to off-nebula area ratio, 2) replace the arf/rmf files of the off-nebula background to those of the on-nebula region for each chip type, and 3) use the \textit{wdata} command to export the resultant count rate spectral model. This exported model multiplied by the exposure then gives the X-ray background contribution to the on-nebula spectrum.  We add the contribution to the non-X-ray background spectrum, using the \textit{ftool} routine \textit{addspec}, to form the total background contribution to be subtracted in the modeling of the on-nebula spectrum of the region  (as demonstrated in Fig.~\ref{f:compare}). 

\begin{table}
\caption{Spectral analysis results of the X-ray background}
\label{t:spec_back}
\begin{tabular}{lr}
\hline
$kT_{1}(\rm{keV})$ & $0.09(0.08-0.10)$ \\
$\rm{EM_{1}(10^{59}cm^{-3})}$ & $7.7(7.2-8.4)$ \\
$N_{\rm H} (10^{22}\rm{cm}^{-2})$ & $0.66(0.64-0.68)$ \\
$kT_{2}(\rm{keV})$ & $0.23(0.22-0.24)$ \\
Abundance (solar)& $5.0$  \\
$\rm{EM_{2}(10^{59}cm^{-3})}$ & $8(7-9)$ \\
$\alpha$ (power-law index)& $2.0(1.9-2.1)$ \\
$\rm{Norm^{pow}}$ $(10^{-3}$$\rm{keV^{-1}}$ $\rm{cm^{-2}}$ $\rm{s^{-1}}$)& $2.2(1.9-2.6)$  \\
$\rm{F^{obs}}$ $(10^{-11}$ ergs $\rm{cm^{-2}}$ $\rm{s^{-1})}$ & $1.49(1.45-1.53)$
\\ 
$\chi^2/dof$ & $\frac{941}{689}=1.37$ \\
\hline
\end{tabular}

\end{table}

\section{Estimation of the PSF contamination in the \suzaku\ spectral analysis} 
\label{a:psf}

\begin{figure}
\includegraphics[width=3.2in,trim={-30 0 30 0}]{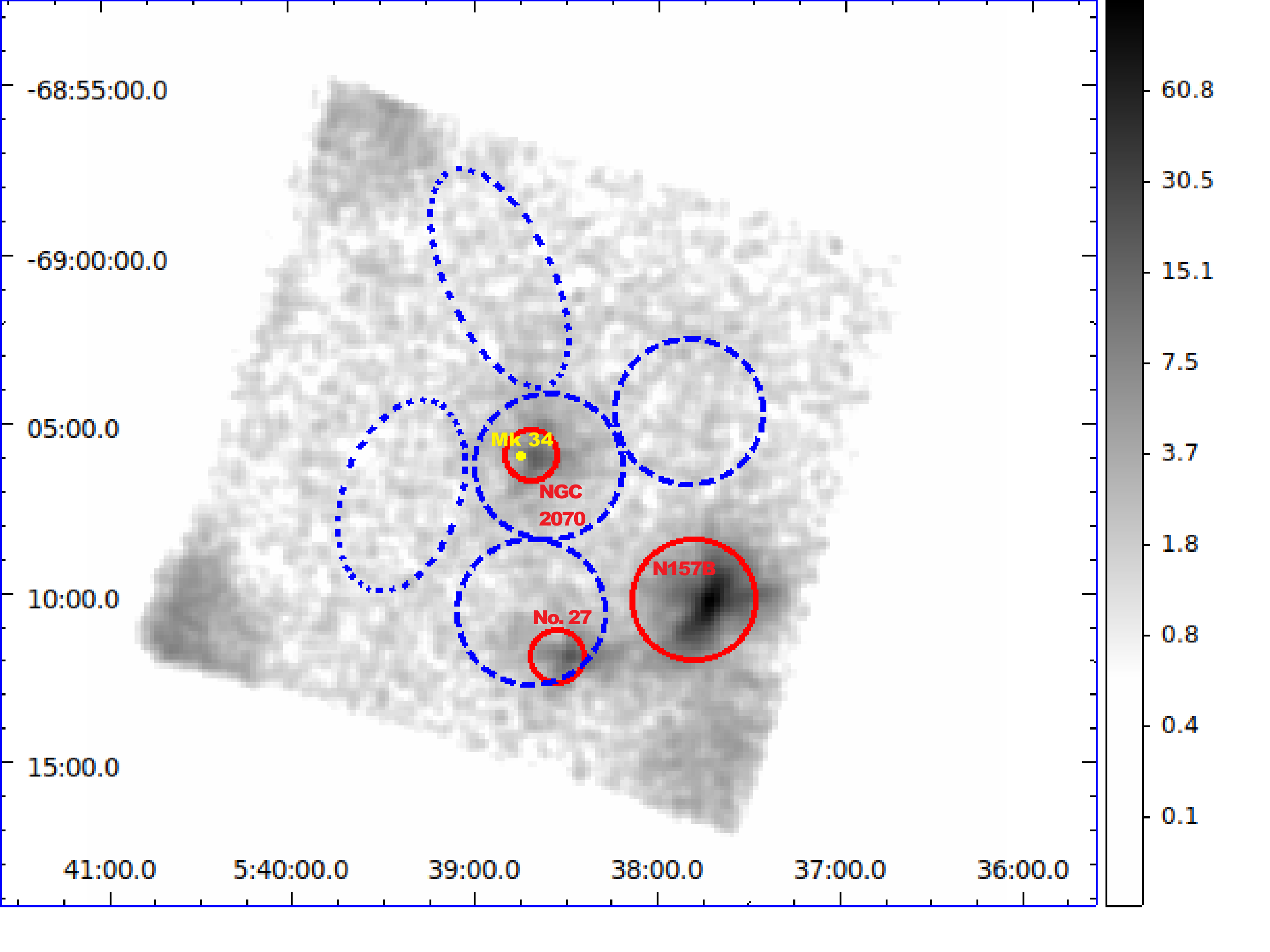}
\caption{The hard-band (4-8~keV) \suzaku\ image with main bright point-like sources marked in red circles. Melnick 34 is marked in a yellow dot. In order to avoid degradation of the data quality, the three corners severely contaminated by calibration sources (as seen in the image) are excluded from the spectral analysis.}
\label{f:point}
\end{figure}

We here estimate the (4-8 keV) hard-band PSF contamination from discrete sources in various \suzaku\ spectral extraction regions in \xs. 
An examination of the hard band \suzaku\ image (Fig. \ref{f:point}) shows three relevant sources: 1) NGC 2070, centered at the NED position R.A., Dec. (J2000) $= 05^h 38^m 42^s, -69^{\circ}06^{\prime}03^{\prime\prime}$; 2) the PWN N157B (SNR J0537.8-6910) at $05^h 37^m 49^s, -69^{\circ}10^{\prime}19^{\prime\prime}$ \footnote{$https://hea-www.harvard.edu/ChandraSNR/snrcat\_lmc.html$}; and 3) a point-like source  (No. 27 in \citet{Townsley2}) at the south-west corner of Region 2 or $05^h 38^m 33^s, -69^{\circ}11^{\prime}59^{\prime\prime}$.
At the \suzaku\ spatial resolution, these sources may all be reasonably assumed to be point-like. 

We use the enclosed energy fraction (EEF) of the PSF as a function of off-source distance (\citealt{Serlemitsos}) to estimate the contamination in various off-center regions used to extracted the spectra (\S~\ref{ss:off-center}). For each region, we sum the contamination from the sources, using their count rates obtained in the 4-8 keV image (Fig. \ref{f:point}). In the total off-center region, for example, the summed contamination is $\sim 5.3 \times 10^{-4} {\rm~counts~s^{-1}}$, compared to the total background-subtracted count rate $\sim 7.0 \times 10^{-3} {\rm~ counts~s^{-1}}$, or 8\% contamination in the 4-8 keV spectrum. The largest contamination fraction, 16\%, occurs in Region 2, chiefly due to the source No. 27 in \citet{Townsley2}. Therefore, we conclude that the PSF contamination from the three identified hard-band sources is at most marginally significant and may not explain the apparent NTX emission in the off-center regions of \xs.  

\section{X-ray absorption model with a log-normal column density distribution}
\label{a:impl-lnnh}
We define a multiplicative X-ray absorption model with a log-normal column density distribution, {\sl lnwabs}, as 
\begin{equation}
\phi(\varepsilon) = \int {\rm exp(}-\sigma_{\varepsilon} N_{H})G(x| \bar{x},\sigma_{x}) dx,
 \label{e:impl-lnnh}
\end{equation}
where $\sigma_{\varepsilon}$ is the X-ray absorption cross section as a function of photon energy $\varepsilon$, $N_{H}$ is the absorbing gas column density, and $G(x| \bar{x},\sigma_{x})$ is the log-normal distribution of $N_{H}$, or
\begin{equation}
  G(x|\bar{x},\sigma_x)=\dfrac{1}{\sqrt{2\pi } \sigma_x} {\rm exp}\Big[-\dfrac{(x - \bar{x})^2}{2\sigma_x^2}\Big],
  \label{e:log-nh}
\end{equation}
in which  $x \equiv {\rm ln} N_{H}$,  while $\bar{x}$ and  $\sigma_x$ are the mean and dispersion of $x$.
The calculated $\phi(\varepsilon) $ can then be multiplied to any model spectrum.

\section{X-ray emission spectral model of optically-thin thermal plasma with a log-normal temperature distribution}
\label{a:impl}
There are existing spectral models for multi-temperature, optically-thin, thermal plasma in XSPEC. The most relevant is the model called {\sl vgadem}.In this model, the differential emission measure (EM) of plasma follows the Gaussian distribution.
An examination of the source code of the model (vgaussDem.cxx) shows that one of the model parameters, {\sl Tsigma}, is actually $\sqrt{2} \sigma$ in the
standard definition of the Gaussian (or normal)  probability distribution:
\begin{equation}
  G(T|\bar{T},\sigma)=\dfrac{1}{\sqrt{2\pi } \sigma} {\rm exp}\Big[-\dfrac{(T - \bar{T})^2}{2\sigma^2}\Big],
  \label{e:gau-T}
\end{equation}
where $\bar{T}$ and $\sigma$ are the mean and dispersion of the temperature variable $T$. Therefore, one needs to be careful in the interpretation of the parameter {\sl Tsigma}. 
With the distribution, one can mathematically calculate the synthesis spectrum ($f_\varepsilon$) of the plasma as 
\begin{equation}
\begin{split}
f_\varepsilon & = K \int \Lambda_\varepsilon G(T|\bar{T},\sigma) dT,
\end{split}
\label{e:spec}
\end{equation}
in which the XSPEC {\sl apec} model definition of the spectral emissivity ($\Lambda_\varepsilon $) is used and  
\begin{equation}
K=\dfrac{10^{-14}}{4\pi D^2} \int n_e n_H dV, 
\label{e:snorm}
\end{equation}
where $n_e$ and $n_H$ are the electron and proton number densities,  $D$ is the distance to the source,
and the integration is over its occupied volume.  In practice, the code approximately constructs the spectrum in a discrete fashion. Spectra are first calculated for a grid of  21 sample temperatures equally spaced within the $-3{\sl Tsigma}$ to $3{\sl Tsigma}$ range around a given $\bar{T}$. Of course, the lower boundary needs to be set so that it is greater than zero; in general, the Gaussian distribution is applicable only when {\sl Tsigma} is very small, compared to $\bar{T}$. These spectra are then summed with the weights corresponding to the Gaussian integrations over the 21 
temperature internals (i.e., the error function differences between the
higher and lower boundaries of the intervals). The sum of all these weights is almost 2 (because the error function ranges from -1 to 1).  

We implement our log-normal temperature distribution model, {\sl vlntd}, by simply replacing $T$ with $x \equiv {\rm ln} T$,  $\bar{T}$ with  $\bar{x}$, and  $\sigma$ with $\sigma_x$ (the dispersion of $x$) in {\sl vgadem}. Thus we have
\begin{equation}  G(x|\bar{x},\sigma_x)=\dfrac{1}{\sqrt{2\pi } \sigma_x} {\rm exp}\Big[-\dfrac{(x - \bar{x})^2}{2\sigma_x^2}\Big].
  \label{e:log-T}
\end{equation}

The construction of the synthesis spectrum is formally same as in Eq.~\ref{e:spec}, except for the change from $G(T|\bar{T},\sigma) dT$ to $G(x|\bar{x},\sigma_x) dx$: i.e.,
\begin{equation}
\begin{split}
f_\varepsilon & = \dfrac{10^{-14}}{4\pi D^2} \int \Lambda_\varepsilon G(x|\bar{x},\sigma_x) dEM\\
& = K \int \Lambda_\varepsilon G(x|\bar{x},\sigma_x) dx,
\end{split}
\label{e:spec-x}
\end{equation}
where $dEM = n_e n_H dV$.
 We choose our fitting parameters to be  $\bar{x}$,  $\sigma_x$, and $K$, in addition to those needed in calculating $\Lambda_\varepsilon$ (including the elemental abundances of the plasma), assuming the collisional ionization equilibrium. 

{\sl vlntd} is a generalization of the single-temperature or Gaussian $EM$ distribution model. 
When $\sigma_x$ is small and the dependence of $x$ on $T$ can be linearized around its mean, it is easy to show that Eq.~\ref{e:log-T} becomes the Gaussian distribution (Eq.~\ref{e:gau-T}). Or when $\sigma_x \rightarrow 0$, the plasma is isothermal. 

The physical meaning of $K$ in Eq.~\ref{e:spec-x}, 
as in Eq.~\ref{e:spec},  is generally not clear. One might interpret it as being the same as the normalization of the 1-T {\sl apec} model~\citep[e.g.][]{deplaa,Mao}. 
However, this interpretation would have to assume that the densities, $n_e$ and $n_H$ or EM, are independent of the temperature so they can be moved out of the integration in Eq.~\ref{e:spec}. This kind of 'isochoric' assumption is typically not applicable to hot plasma in superbubbles. So alternative interpretation is desirable.

We interpret the log-normal temperature distribution differently, which leads to a more realistic and general connection of $K$ to the physical state of the plasma.  Our interpretation is based on the assumption that the differential volume occupation  of the plasma as a function of $x$ ($\equiv {\rm ln} T$) is 
\begin{equation}
\dfrac{dV}{dx} = V_t G(x|\bar{x}_V,\sigma_x),
\label{e:dv}
\end{equation}
where $V_t$ is the total volume of the plasma from which the spectral data is extracted, while $ G(x|\bar{x}_V,\sigma_x)$ is the same  as in Eq.~\ref{e:log-T}, except for the subscript $V$ of $\bar{x}_V$, which means that the average here is over the volume (instead of the EM).

To make the connection to the EM distribution, we need to consider the equation of state of the plasma. Assuming ideal gas for the plasma, we have the pressure $P_{th}=nk_BT$. Here we specifically consider the isobaric (constant pressure) case, as commonly assumed in existing studies. Then the total particle number density $n$ ($\equiv n_e+n_i$) can be expressed as $n = \eta (n_en_H)^{1/2}$, where $\eta = \frac{(\mu_H \mu_e)^{1/2}}{\mu}$, while $\mu$, $\mu_H$, and $\mu_e$ are the mean molecular weights of the total particle, hydrogen, and electron, respectively. $\eta \approx 2.1$ for a typical metallicity expected for hot plasma in superbubbles. 
Neglecting any molecular weight variation with the temperature, we have $n \propto 1/T$. With these, we can now express the differential distribution of the EM as 
\begin{equation}
\begin{split}
\dfrac{d EM}{dx} & \equiv n_en_H \dfrac{dV}{dx} \\
& =\dfrac{P_{th}^2 V_t}{\eta^2 k_B^2} \dfrac{G(x|\bar{x}_V,\sigma_x)}{T^2} \\
& =\dfrac{P_{th}^2 V_t}{\eta^2 k_B^2} e^{-2(\bar{x}_V - \sigma_x^2)} G(x|\bar{x}_V-2\sigma_x^2,\sigma_x).\\
\label{e:em0}
\end{split}
\end{equation}
With $\bar{x} \equiv \bar{x}_V-2\sigma_x^2$, which is the EM-weighted mean of 
$x$ (in contrast to the volume-weighted mean, $\bar{x}_V$), 
the above equation can  be expressed as
\begin{equation}
\dfrac{d EM}{dx} = \dfrac{P_{th}^2 V_t}{\eta^2 k_B^2} e^{-2(\bar{x} + \sigma_x^2)} G(x|\bar{x},\sigma_x).
\label{e:dem}
\end{equation}
Therefore, the differential distribution of the EM as a function of $T$ is log-normal as well, which 
generally holds  for any power law dependence of $n$ on $T$ (e.g.,  for a polytrope equation of state). In such a case, we can then apply Eq.~\ref{e:log-T} in the X-ray spectral modeling.

Placing Eq.~\ref{e:dem} in Eq.~\ref{e:spec-x}, we can then obtain
\begin{equation}
K  = \dfrac{10^{-14}}{4\pi D^2}\dfrac{P_{th}^2 V_t}{\eta^2 k_B^2} e^{-2(\bar{x} + \sigma_x^2)}.
\label{e:norm}
\end{equation}
When $\sigma_x \rightarrow 0$, $K$ recovers to the XSPEC model normalization of isothermal plasma  (e.g., {\sl apec}):
\begin{equation}
K  \rightarrow \dfrac{10^{-14}}{4\pi D^2}V_t n_en_H.
\label{e:norm2}
\end{equation}

In general with $\bar{x}$, $\sigma_x$, and $K$ obtained from spectral fitting, we can infer the pressure, for example, as
\begin{equation}
\begin{split}
P_{th} & = \Big[\dfrac{4\pi D^2}{10^{-14}}\dfrac{\eta^2 K}{V_t}\Big]^{1/2} k_B e^{\bar{x}+\sigma_x^2}.
\end{split}
\label{e:p}
\end{equation}
Notice that $\bar{T} \equiv e^{\bar{x}}$ is a factor of $e^{2\sigma_x^2}$ smaller than $\bar{T}_V \equiv e^{\bar{x}_V}$.  Clearly, the pressure  in the present case is a factor of $e^{\sigma_x^2}$ greater than the estimate assuming that the plasma is isothermal (i.e., $\sigma_x =0$). These parameter biases are due to the fact that the emission is proportional to the EM. 

\section{estimation of physical parameters of the plasma}
\label{a:para}

To infer the mass, thermal energy and pressure of the plasma, we start from the estimation of its occupied volume. A uniform sphere is assumed for Region 1, 2 and 3, while for Region 4 and 5, an ellipsoidal volume is assumed for both prolate and oblate cases. For the entire nebula, we roughly estimate it to be the sum of a southeast-northwest oriented bubble and two protrusions: one to the southwest and the other to the northeast (Fig.~\ref{f:im_reg}B). The bubble is regarded as a 104 pc $\times$ 50 pc ellipsoid, and the two protrusions are taken as two half ellipsoidal caps of a 36 pc $\times$ 138 pc ellipsoid. Similar to individual regions, both prolate and oblate cases are considered. The dimensions and estimated volumes of each region and the entire nebula are listed in Table~\ref{t:para_derived}. The lower and upper limits to the total volume $V_T$ are calculated as (145 and 363) $\times 10^{4} {\rm~pc^{3}}$.

For the log-normal temperature distribution model, as detailed in Appendix \ref{a:impl}, we infer various physical parameters of the plasma. By integrating Eq.~\ref{e:em0} over $x$, we obtain the total EM as
\begin{equation}
EM  =\dfrac{P_{th}^2 V_t}{\eta^2 k_B^2} e^{-2(\bar{x} + \sigma_x^2)}.
\label{e:em}
\end{equation}
We estimate the total plasma mass in $V_{t}$ as
\begin{equation}
\begin{split}
M & = \int \rho dV\\ 
& = \int \dfrac{P_{th}\mu m_{p} V_{t}}{k_B T} G(x|\bar{x}_V,\sigma_x) dx \\
& = \dfrac{P_{th}\mu m_{p} V_{t}}{k_B} e^{-\bar{x}-\sigma_x^2/2},
\label{e:mass}
\end{split}
\end{equation}
where $m_{p}$ is the proton mass and $P_{th}$ can be obtained from Eq. \ref{e:p}. Compared with the isothermal case, the mass is dropped by a factor of $e^{-\sigma_x^2/2}$.  Finally, the total thermal plasma energy is simply $E_{th} =\frac{3}{2} P_{th} V _{t} \sim (2.75-4.35) \times 10^{52} f_h^{1/2}
{\rm~ergs}$, where $f_{h} \sim 1$ is the effective filling factor of the plasma in the volume $V_{t}$ (Table~\ref{t:para_derived}).

\end{document}